\documentclass[final]{svjour3}
\usepackage{amsmath,amssymb}
\usepackage[usenames,dvipsnames]{color}
\usepackage[pdftex]{graphicx}
\usepackage[normalem]{ulem}

\definecolor{darkred}{rgb}{0.7,0,0}
\definecolor{darkblue}{rgb}{0,0,0.5}
\definecolor{newtext}{rgb}{0.3,0,0}
\newcommand{\arccot}{\mathop{\rm arccot}\nolimits}
\newcommand{\hyphen}{\mathbin{\text{-}}}

\newcommand{\rem}[1]{{\sout{#1}}}

\newcommand{\url}[1]{{\tt #1}}

\journalname{General Relativity and Gravitation}
\begin{document}

\title{QND measurements for future gravitational-wave detectors}

\author{
Yanbei Chen
\and
Stefan~L.~Danilishin
\and
Farid~Ya.~Khalili
\and
Helge M\"uller-Ebhardt}

\institute{
Yanbei Chen
\at
Theoretical Astrophysics 350-17, California Institute of Technology, Pasadena, California 91125, USA
\and
Stefan L. Danilishin
\at
Physics Faculty, Moscow State University, Moscow 119992, Russia
\and
Farid Ya. Khalili
\at
Physics Faculty, Moscow State University, Moscow 119992, Russia
\and
Helge M\"uller-Ebhardt
\at
Max-Planck-Institut f\"ur Gravitationsphysik (Albert-Einstein-Institut) and Universit\"at Hannover, Callinstr.~38, 30167 Hannover, Germany
}

\maketitle

\begin{abstract}

Second-generation interferometric gravitational-wave detectors will be operating at the Standard Quantum Limit, a sensitivity limitation set by the trade off between measurement accuracy and quantum back action, which is governed by the Heisenberg Uncertainty Principle.  We review several schemes that allows the quantum noise of interferometers to surpass the Standard Quantum Limit significantly over a broad frequency band. Such schemes may be an important component of the design of  third-generation detectors.

\end{abstract}


\section{Introduction}

The first generation large-scale laser interferometric gravitational-wave (GW) detectors (LIGO \cite{Abramovici1992,LIGOsite}, Virgo \cite{Ando2001,VIRGOsite}, GEO600 \cite{Willke2002,GEOsite}, and TAMA \cite{TAMAsite}) are either on duty now or have finished the first stage of their life cycle and are in preparation to an upgrade. Second-generation detectors, e.g., Advanced LIGO, Advanced Virgo, GEO-HF and LCGT aim at increasing sensitivity by about one order of magnitude. This is archived by means of quantitative improvements (higher test-mass weight, higher optical power, improved quality of the technical components such as the test-mass suspension and the mirror and beam splitter materials, smaller losses) and evolutionary changes of the interferometer configurations, most notably, by introduction of the signal recycling mirror~\cite{Thorne2000,Fritschel2002,Smith2009,Acernese2006-2,Willke2006,LCGTsite}.

As a result, the second-generation detectors will be \textit{quantum noise limited}. At higher GW frequencies, the main sensitivity limitation will be due to phase fluctuations of the light inside the interferometer (shot noise). At lower frequencies, the random force created by the amplitude fluctuations (radiation-pressure noise) will be the main or among the major contrubutors into the sum noise. The best sensitivity point, where these two noise sources become equal, is known as the Standard Quantum Limit (SQL) \cite{92BookBrKh}, which, in a broader context, characterizes the regime in which the quantum measurement noise (shot noise) becomes equal to the back action noise (radiation pressure noise) -  the latter one being a consequence of the Heisenberg uncertainty principle (see details in Sec.\,\ref{sec:SQL}).

Topologies of third-generation detectors are far from being established. For example, a triangular or a rectangular shape of the vacuum system for the  {\it Einstein Telescope} (ET) \cite{Freise2009}. It could accommodate several independent Michelson interferometers, but evidently can be used also more ``synergetically'' for non-Michelson topologies.  If third-generation detectors are to improve sensitivities by yet another order of magnitude, the main noise sources we face would be: (i) the low-frequency portion of quantum noise, (ii) thermal noises from suspension and mirror coatings, (iii) Newtonian gravity gradient and vibrational seismic noises, and (iv) shot noise, where (i) -- (iii) are mostly in low frequencies (below 100\,Hz), and (iv) is for high frequencies.  Methods for tackling these noises are {\it orthogonal}: (i) argues for heavy mirrors, lower optical power, (ii) argues for mirrors with large radii, good mechanical properties and low absorption, lower optical power and low temperature, (iii) requires careful design of the suspension system and noise cancellation by monitoring ground-motion, and (iv) argues for high optical power. Aside from increasing the arm length of the detector (which improves sensitivity in the entire band but incurs huge cost), we may  have to consider separate detectors for coverage of the entire ground-based frequency window of 1\,Hz up to 10\,kHz.

This paper focuses on designs of interferometers that beat the SQL significantly in a broad frequency band, often named Quantum Non-Demolition (or QND) schemes.~\footnote{Initially (see \cite{77a1eBrKhVo}) this term was proposed for the specific class of quantum measurements where the measured observable commutes with the unperturbed Hamiltonian and thus is not ``demolished'' during the measurement (for example, energy of the oscillator non-linearly coupled with the meter). In this case, monitoring of this observable with arbitrary high precision is possible, allowing to detect arbitrary small external perturbation. Today the term QND is frequently (and, in our opinion, imprecisely) applied to any mechanical measurement which can overcome the free mass SQL within a certain frequency range. The three schemes considered in this paper, strictly speaking, do not comply with the initial QND definition. However, they represent the closest approximation to it currently available.} In particular, we mainly consider three schemes: (i) the {\it variational readout scheme,} in which frequency-dependent homodyne detection is implemented by means of additional km-scale filter cavities, and allows broad-band back-action evasion for a broad-range of interferometers (Sec~\ref{sec:positionmeter}), (ii) the {\it speed meter}, which measures speed of test masses, and can easily evade back-action noise without broad-band frequency-dependent filtering (Sec.~\ref{sec:speedmeter}), and (iii)  the {\it intracavity readout scheme}, in which effect of the GW is first transduced onto the motion of a mirror in its local inertial frame, via {\it optomechanical coupling}, and then detected by measuring this local motion (Sec.~\ref{sec:intracavity}).

It is sometimes argued that instead of overcoming the SQL it is possible just to reduce its value by using heavier test masses which have naturally lower radiation-pressure noise. However, we believe that QND techniques are still very useful because: (i) the SQL scales with $1/\sqrt{M}$, and therefore significantly heavier mirrors will be needed, and they bring newer technical issues, (ii) conventional interferomter (will be defined explicitly in the next section) has a noise spectrum that touches the SQL at one frequency, and then increases sharply as $f^{-2}$ at lower frequencies, which makes QND schemes useful even for such heavy mirrors.  While it is most probable that QND schemes would be useful for low-frequency detectors, there is still indication that configurations arising from the study of QND interferometers could improve high-frequency sensitivities.

Throughout our review, we will assume that frequency-independent squeezed vacuum (which can serve for improving the interferometers' sensitivity) is available for injection into the dark port of the interferometer.  For the sake of formulae simplicity we confine ourselves to the particular case of phase quadrature squeezed light that is necessary to reduce the photon shot noise dominating at high signal frequencies, as it was first proposed by Caves~\cite{Caves1981}. More detailed analysis exceeds this paper frame and can be found, for example, in the articles \cite{02a1KiLeMaThVy,Harms2003}.

Main notations used throughout the paper are listed in Table~\ref{tab:notations}.

\begin{table}[t]
  \begin{tabular}{|c|l|c|l|}
    \hline
      Quantity & Description &       Quantity & Description \\
    \hline
      $c$         & Speed of light   &       $T$         & Input mirror transmissivity \\
      $\hbar$     & Plank's constant    &       $\gamma$    & interferometer's 1/2-bandwidth  \\
      $\omega_o$  & Optical pumping frequency &       $I_c$       & Circulating power \\
      $\Omega$    & Signal (side-band) frequency &       $e^{-2r}$    & Power squeeze factor  \\
      $L$         & Arm length  &        $\zeta$     & Homodyne angle \\
      $M$         & Mirror mass &       $\eta$      & Unified quantum efficiency \\
    \hline
  \end{tabular}
  \caption{Main notations used in this paper.}\label{tab:notations}
\end{table}

\section{Standard Quantum Limit}\label{sec:SQL}

As a starting point, we consider a Michelson interferometer with Fabry-Perot cavities in its arms (as a {\it position meter}, in the sense that below the cavity linewidth, the output phase modulation is directly proportional to the position of mirrors), which is the standard design for the first and second generation detectors. We assume also that the pump laser frequency is equal to the eigenfrequency of the interferometer anti-symmetric (signal) mode. This means that either no signal recycling is used at all, or both the arm cavities and the signal recycling cavity are tuned in resonance \cite{Buonanno2003}.\footnote{In principle, the variant with non-zero opposite-sign detunings of the arm cavities and signal recycling cavity is also possible, while it can hardly be  considered practical, as it provides the same result but for higher (experimental) ``price''.}  We shall refer to this scheme, where homodyne readout of the outgoing light phase quadrature (phase measurement) is implied, as a {\it conventional interferometer}.

The first treatment of quantum noise in laser interferometric GW detectors was done by Caves \cite{Caves1981}. He pointed out the existence of two above mentioned kinds of quantum noise (shot noise and radiation pressure noise), that together govern the sensitivity of such detectors. He has also calculated the measurement uncertainties, originating from these two kinds of quantum noise (see Eqs.\,(2.36) and (2.20) of that paper). Another comprehensive consideration of quantum noise in GW interferometers, based on the earlier works by Caves, was performed in Section II of Ref.~\cite{02a1KiLeMaThVy}. In this paper, quantum fluctuations of light coming out the dark port of the interferometer and carrying also GW signal, are expressed by means of input-output relations in terms of fluctuations of the vacuum field entering the interferometer from the dark port (see Eqs.~(16)). Eq.~(18) of this article introduces the quantitative measure of coupling strength $\mathcal{K}$ between the optical fields and mechanical motion of the mirrors, that governs both the interferometer response to gravitational wave and to radiation pressure fluctuations. This concept turned out to be extremely useful, and the sum quantum noise of conventional interferometer one can write in terms of this coupling constant as:
\begin{equation}\label{S_SQL}
  S^h_{\rm PM} = \frac{h^2_{\rm SQL}}{2}
    \left[\frac{e^{-2r}}{\mathcal{K}_{\rm PM}(\Omega)} + \mathcal{K}_{\rm PM}(\Omega)e^{2r}\right] .
\end{equation}
Here $S^h$ is the ``one-sided'' (standard for the GW community) spectral density of the sum quantum noise written in units of equivalent fluctuations of GW strain $h(t)$, ``PM'' stands for ``position meter'', and
\begin{equation}\label{eq:SQL}
  h_{\rm SQL}^2 = \frac{8\hbar}{M\Omega^2L^2}
\end{equation}
is the minimal value of $S^h_{\rm PM}$, known as the spectral form of the SQL (here and below, we  suppress the $\Omega$ dependence of $S^h$ and $h_{\rm SQL}$). We define the opto-mechanical coupling constant in the same manner as in Ref.~\cite{02a1KiLeMaThVy}:
\begin{equation}\label{KPM}
  {\cal K}_{\rm PM}(\Omega) = \frac{2I_c}{I_c^{\rm SQL}}\,
    \frac{\gamma^4}{\Omega^2(\gamma^2 + \Omega^2)}
\end{equation}
with $\gamma$ being the half-bandwidth of the arm cavity,
\begin{equation}\label{I_c_PM}
  I_c =  {4\,I_{\rm BS}}/{T}
\end{equation}
is the total optical power circulating in both arms of the interferometer, $I_{\rm BS}$ is the {\it input power} at the beamsplitter, $T$ is the power transmissivity of the arm cavities input mirrors, and
\begin{equation}\label{I_c_SQL}
  I_c^{\rm SQL} = \frac{McL\gamma^3}{4\,\omega_o} \,.
\end{equation}

The first term in Eq.\,\eqref{S_SQL} owes to shot noise while the second term represents the contribution by radiation-pressure noise. One can immediately see the important feature of these two terms noted by Caves \cite{Caves1981}, namely their opposite dependence on optical power. This implies the existence of a minimal noise at a given frequency $\Omega$ that can be achieved if the circulating optical power takes the optimal value:
\begin{equation}
  I_c^{\rm opt} = \frac{\Omega^2(\gamma^2 + \Omega^2)}{\gamma^4}\,I_c^{\rm SQL}e^{-2r} \,.
\end{equation}
The corresponding minimum value known as the SQL (\ref{eq:SQL})
is reached when the shot noise contribution to (\ref{S_SQL})
equals to the contribution of the radiation pressure noise.

Putting typical numbers for contemporary laser GW detectors, with
$M\gtrsim10\,{\rm kg}$, $L\gtrsim1\,{\rm Km}$, and
$\Omega\sim\gamma\sim10^3$, one sees that this power can be of the
order of megawatts. This is a potential source of significant
difficulties, such as thermal lensing or parametric instabilities.
This fact motivated Caves and Unruh
\cite{Caves1981,Unruh1982} to propose the use of (phase) squeezed
light ($r>0$) in order to decrease the shot noise (and in crease
proportionally radiation pressure noise), and reduce, as a result,
the optical power necessary to reach the SQL.

\section{Variational Readout Schemes}\label{sec:positionmeter}

\subsection{Quantum noises correlation}\label{ssec:variational}

In the previous section, an important resource was omitted, namely
possible cross-correlation of the shot noise and the radiation
pressure noise as it has first been realized by Unruh
\cite{Unruh1982}. Following this initial paper \cite{Unruh1982},
several authors
\cite{87a1eKh,JaekelReynaud1990,Pace1993,96a2eVyMa,02a1KiLeMaThVy,Arcizet2006}
developed the technique, that allowed to detect the action of a
weak classical force on a test mass with precision not limited by
the SQL, utilizing the cross-correlation between the shot noise
and radiation pressure noise for subtraction of the latter from
the output signal.

Simple frequency-independent correlation can be introduced by
measuring instead of phase quadrature some arbitrary quadrature
component of the outgoing light via balanced homodyne
detection, defined by the homodyne angle $\zeta$ (the case of
arbitrary $\zeta$ was considered in \cite{Caves1981}, but the
cross-correlation was not taken into account). The detailed
analysis of Fabry-Perot Michelson position meters, with account
for quantum noise correlation, was done by Kimble {\it et al} in
Ref.\,\cite{02a1KiLeMaThVy}. It immediately follows from Eq.\,(56)
of that paper, that in the ideal lossless case ($\eta=1$), the
quantum-noise spectral density of a position meter with arbitrary
homodyne detection angle $\zeta$ is equal to
\begin{equation}\label{S_PM}
  S^h_{\rm PM} = \frac{h^2_{\rm SQL}}{2}
    \frac{e^{-2r} + [\cot\zeta - \mathcal{K}_{\rm PM}(\Omega)]^2e^{2r}}{\mathcal{K}_{\rm PM}(\Omega)} \,.
\end{equation}
It is easy to see from this equation, that at any given frequency $\Omega$, the back action term (proportional to $\mathcal{K}_{\rm PM}$) can be canceled by setting
\begin{equation}\label{zPM}
  \zeta = \arccot {\cal K}_{\rm PM}(\Omega) \,,
\end{equation}
that yields
\begin{equation}\label{S_PM_VM}
  S^h_{\rm PM} = \frac{h^2_{\rm SQL}}{2}\,\frac{e^{-2r}}{\mathcal{K}_{\rm PM}(\Omega)} \,.
\end{equation}
This spectral density contains shot noise only, and can in
principle beat the SQL by an arbitrary amount, given a large
enough circulating power $I_c$ and/or squeezing factor
$e^{2r}$.

Further development of this idea, providing tools for overcoming
the SQL in wide frequency band  and called \textit{variational
measurement}, was proposed in Ref.\,\cite{02a1KiLeMaThVy}. Therein
the light, leaving the main interferometer, passes through two
additional Fabry-P\'erot filter cavities ({\it Kimble filters}),
before it is measured by conventional homodyne detector. Those
cavities role is to introduce frequency-dependent phase shift,
thus implementing measurement of a frequency-dependent quadrature,
which obeys Eq.\,\eqref{zPM} in broad band. Filter cavities should
be detuned from the pumping laser frequency in an optimal way that
is being discussed in Appendix C of Ref.\,\cite{02a1KiLeMaThVy}.

More generally, as shown in Appendix A of Ref.~\cite{Purdue2002} and in \cite{Buonanno2004},  any rotation of the form
\begin{equation}
\cot\zeta(\Omega) = \frac{\sum_{j=1}^{n} a_j \Omega^{2j}}{\sum_{j=1}^{n} b_j\Omega^{2j}}\,, \quad
  a_n b_n \neq 0
\end{equation}
can be achieved with $n$ filters. The bandwidths and detuning frequencies of the filters has to be of the same order of magnitude as the detection frequency scale.

\subsection{Optical losses}\label{ssec:losses}

Optical losses play a crucial role in variational readout schemes,
because cancellation of radiation pressure noise is achieved with
decreased signal content in the readout
quadrature~\cite{02a1KiLeMaThVy}. Optical losses can arise from
several origins, e.g.,  lossy mirrors, imperfect mode matching,
finite quantum efficiency of photodetectors. For simplicity, we
use  one effective noise that enters the final output field via an
imaginary beamsplitter that mixes output of an ideal
interferometer with an additional vacuum,  with weights
$\sqrt{\eta}$ and $\sqrt{1-\eta}$, where $\eta$ can be viewed as
an effective quantum efficiency of the photodetector. In general,
$\eta$ depends on $\Omega$. The main source of this dependence on
frequency arise from the filter and (if $r\ne0$) from the arm
cavities. For them,
\begin{equation}
  1-\eta_{\rm cav} \sim {cA^2}/{(4L\gamma)} \sim 10^2\times A^2\times\left({10\,{\rm km}}/{L}\right)\
    \times\left({100\,{\rm s}^{-1}}/{\gamma}\right),
\end{equation}
where $A^2$ is the loss per bounce. Assuming the moderately optimistic value of $A^2\sim10^{-5}$, $1-\eta_{\rm cav}\sim10^{-3}$ is an order of magnitude better than the $1-\eta\gtrsim10^{-2}$ which can be provided by modern photodetectors and the auxiliary output optics. Therefore, we neglect in this section frequency dependence introduced by $\eta_{\rm cav}$.

The total quantum noise spectral density of the lossy interferometer is
\begin{equation}\label{S_PM_loss}
  S^h_{\rm PM} = \frac{h^2_{\rm SQL}}{2}
    \times\frac{
      e^{-2r} + \xi_{\rm loss}^4/\sin^2\zeta
      + [\cot\zeta - \mathcal{K}_{\rm PM}(\Omega)]^2e^{2r}
    }{\mathcal{K}_{\rm PM}(\Omega)} \,,
\end{equation}
where we have introduced a convenient factor
\begin{equation}  \xi_{\rm loss} = \sqrt[4]{(1-\eta)/\eta} \,. \end{equation}
In the conventional case of $r=0$ and $\zeta=\pi/2$, the influence of  optical losses is negligible: the factor $1/\eta$ only increases the shot noise by a few percents.  However, losses limit the gain in the shot noise spectral density, provided by squeezing (and thus the gain in the optical power) by the value $(1-\eta)/\eta$. In the variational readout case, losses become even more important because $\zeta$ tends to be small [Cf.~Eq.~\eqref{KPM} and \eqref{zPM}].  A re-optimization of $\zeta(\Omega)$ in presence of loss leads to
\begin{equation}
  \zeta(\Omega) = \arccot\frac{\mathcal{K}_{\rm PM}(\Omega)}{1 + \xi_{\rm loss}^4e^{-2r}} \,,
\end{equation}
which results in a total noise spectrum of
\begin{equation}\label{S_PM_VM_loss}
  S^h_{\rm PM\,VM} = \frac{h^2_{\rm SQL}}{2}\left[
    \frac{e^{-2r} + \xi_{\rm loss}^4}{{\cal K}_{\rm PM}(\Omega)}
    + \frac{\xi_{\rm loss}^4{\cal K}_{\rm PM}(\Omega)}{1 + \xi_{\rm loss}^4e^{-2r}}
  \right],
\end{equation}
where a back-action term, proportional to $\mathcal{K}_{\rm PM}(\Omega)$, persists, albeit reduced by $\approx 1-\eta$. This presence of back-action leads to
\begin{equation}\label{xi_loss}
  \frac{S^h_{\rm PM\,VM}}{h^2_{\rm SQL}}
    \ge \xi_{\rm loss}^2\sqrt{\frac{e^{-2r} + \xi_{\rm loss}^4}{1 + \xi_{\rm loss}^4e^{-2r}}}
\end{equation}
(see also \cite{02a1KiLeMaThVy}), with equality achievable at the unique frequency at which two terms in the square brackets in (\ref{S_PM_VM_loss}) are equal to each other (similar condition and estimate can be found in the abstract to \cite{02a1KiLeMaThVy}). Thus,
\begin{equation}
  \sqrt{S^h_{\rm PM\,VM}}/h_{\rm SQL} = \xi_{\rm loss}
\end{equation}
is the best SQL-beating factor one can hope to achieve given $\eta$ without squeezing. With sufficiently deep squeezing, $e^{-2r}\ll\xi_{\rm loss}^4$, it is possible to achieve
\begin{equation}
  \sqrt{S^h_{\rm PM\,VM}}/h_{\rm SQL} = \xi_{\rm loss}^2 \,.
\end{equation}
For $\eta=0.95$, which is moderately optimistic for current  photodiodes, the best loss-limited SQL beating  is $\sqrt{S^h}/h_{\rm SQL}\simeq0.5$ without squeezing and $\sqrt{S^h}/h_{\rm SQL}\simeq0.22$ with squeezing.

It is evident from this simple consideration that optical losses can become a real problem for sub-SQL sensitivity schemes that rely on quantum noise reduction based on optimal correlation between measurement and back action noise.

\section{Quantum speed meter}\label{sec:speedmeter}

\subsection{Overview of speed meter topologies}

One way of obtaining a sensitivity better than the SQL is to
measure a QND observable that remains unperturbed by the meter's
back action during a continuous measurement. Momentum $p$ of a
free mass is a QND observable, yet practical ways of directly
measuring the momentum of a free mass has not been invented. The
closest is to measure the speed $v$ of a test mass --- yet as soon
as one couples $v$ to an external observable, the canonical
momentum $p$ becomes different from $mv$ --- and even though the
canonical momentum $p$ remains QND, $mv$ will be perturbed by a
continuous measurement.  Nevertheless, speed meters turn out to
offer broadband beating of the SQL more easily than position
meters. The basic idea of the quantum speed meter was first
proposed by Braginsky and Khalili in \cite{90a1BrKh}.

The general approach to speed measurement is to use pairs of position measurements separated by a time delay $\tau\stackrel{<}{_\sim} 1/\Omega$, where $\Omega$ is the characteristic signal frequency. The successive measurement should ideally be coherent, i.e., they should be performed by  the same photons.
\begin{figure}[t]
 \includegraphics[width=0.38\textwidth]{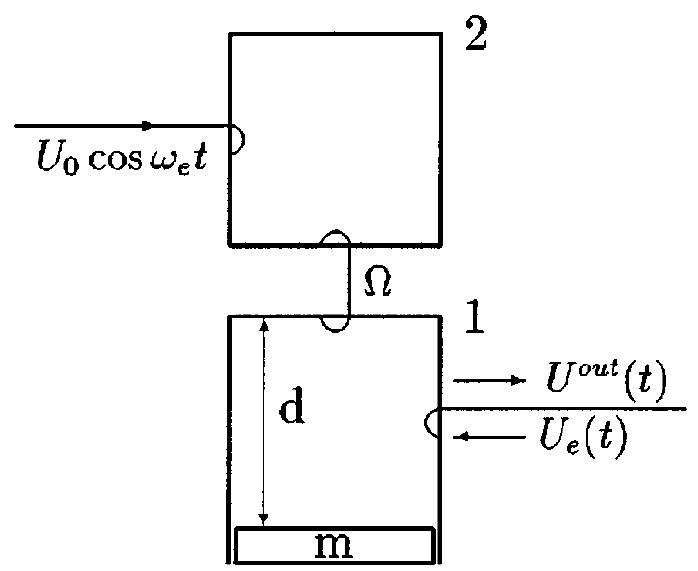}\hfill
 \includegraphics[clip,viewport=0 0 300 265,width=0.58\textwidth]{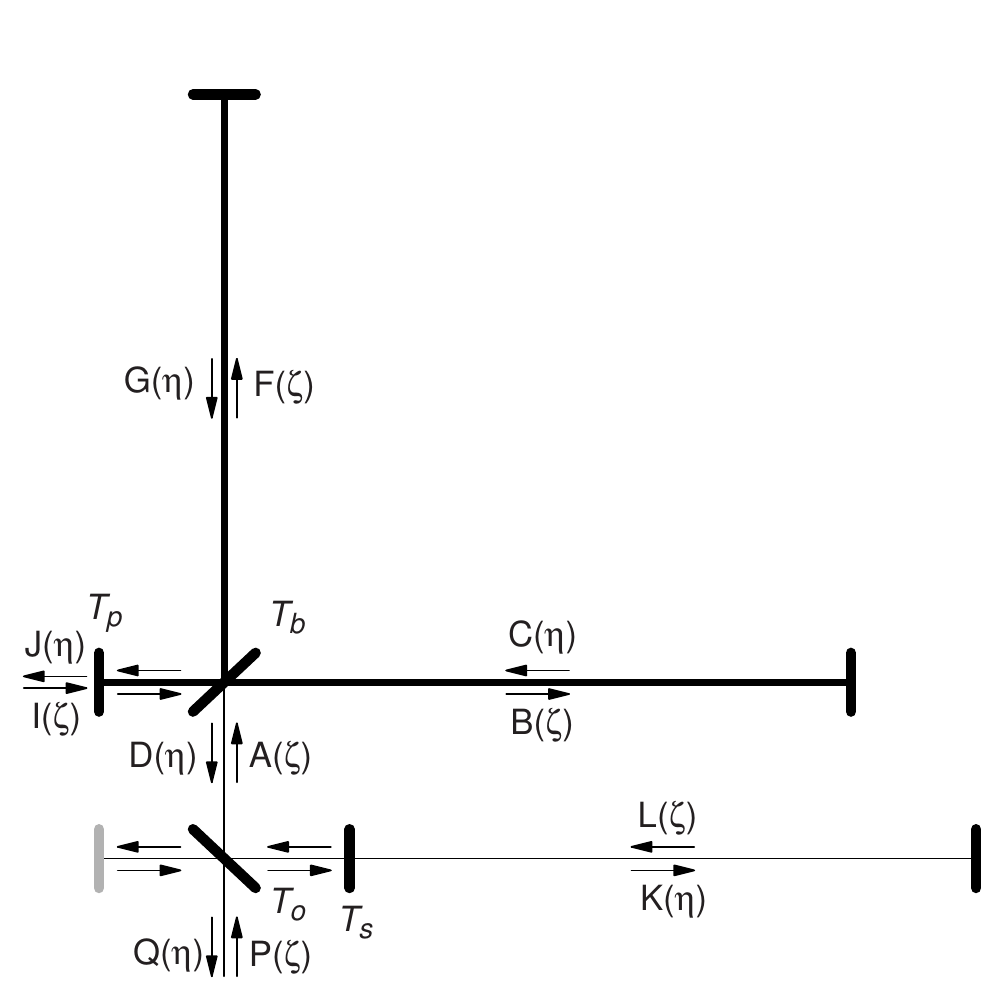}
\caption{Left: schematic diagram of the microwave speed meter on coupled cavities as given in \cite{90a1BrKh,00a1BrGoKhTh}. Right: optical version of coupled cavities speed meter proposed in \cite{Purdue2002}.
  }\label{figSM2}
\end{figure}
In Ref.~\cite{00a1BrGoKhTh}, Braginsky, Gorodetsky, Khalili and Thorne (BGKT) analyzed the  microwave speed meter scheme first proposed in \cite{90a1BrKh} in detail, as a possible realization of local meter for an intra-cavity GW detector that will be considered in Sec.\,\ref{sec:intracavity}.  This scheme uses two identical coupled microwave resonators, as shown in Fig.\,\ref{figSM2} (left), one of which (2) is pumped on resonance though the input waveguide, so that another one (1) becomes excited at its frequency $\omega_e$. The eigenfrequency of resonator~1 is modulated by the position $x$ of the test mass, and puts a voltage signal proportional to position $x$ into resonator~2, and a voltage signal proportional to velocity $dx/dt$ into resonator~1. The velocity signal flows from resonator~1 into an output waveguide, from which it is monitored. One can understand the production of this velocity signal as follows. The coupling between the resonators causes voltage signals to slosh periodically from one resonator to the other at the frequency $\Omega$. After each cycle of sloshing, the sign of the signal is reversed, so the net signal in resonator 1 is proportional to the difference of the position at times $t$ and $t+2\pi/\Omega$, {\it i.e.} is proportional to the test-mass velocity so long as the test mass' frequencies $\omega$ of oscillation are $\omega\ll\Omega$. BGKT also proposed a scheme of an optical speed meter based on the same principle, namely to use two coupled Fabry-P\'erot cavities, with one pumped resonantly, and the second one kept unexcited and serving as a ``sloshing cavity'' (see Fig.\,4 of \cite{00a1BrGoKhTh} for detail). This idea was further elaborated by Purdue, who found that  the BGKT two-cavity scheme, although can beat the SQL broadband, requires huge pumping power, and can only evade classical laser noise if two such two-cavity schemes are used simultaneously~\cite{Purdue2001}. Purdue and Chen~\cite{Purdue2002} subsequently improved upon the BGKT two-cavity scheme, converting the differential mode of a Michelson interferometer into cavity 1, and an additional, km-scale cavity into cavity 2, thus making a practical interferometer configuration, see Fig.\,\ref{figSM2} (right).

Khalili and Levin~\cite{96a1KhLe} proposed another  version of speed-meter-based GW detector, using a Doppler phase shift that appears in light wave after passing through a moving, rigid, impedance-matched Fabry-P\'erot cavity. It was suggested in \cite{96a1KhLe} to attach small rigid Fabry-P\'erot resonators to test masses of GW detector and tune them in resonance when at rest. When the cavity starts moving it sees the incoming light Doppler shifted, and the light emerging from the other side gets phase shifted by $ \delta\varphi = ({\omega_o\tau^*}/{c})\,v$, where $v$ is the cavity velocity, while $\omega_o$ and $\tau^*\sim\gamma^{-1}$ are its eigenfrequency and ringdown time, respectively. Unfortunately, the requirement that the cavity have to be a stiff one puts limitation on the cavities length and thus on the achievable values of the ringdown time $\tau^*$. An estimate show, that due to this limitation, the optical power in this scheme have to be unrealistically high to compensate the small factor $\omega_o\tau^*/c$.
\begin{figure*}[t]
  \includegraphics[width=.48\textwidth]{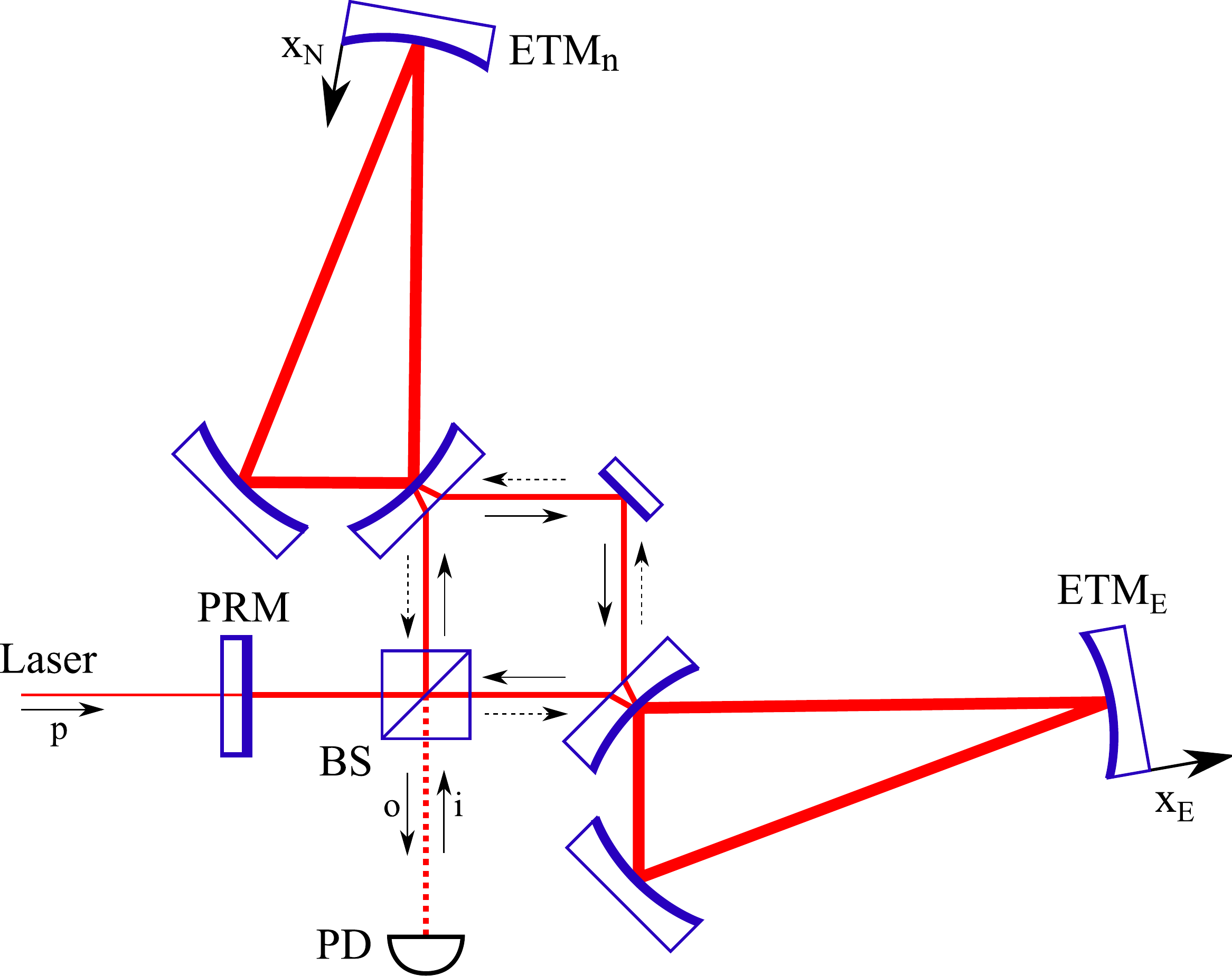}\hfill\includegraphics[width=.48\textwidth]{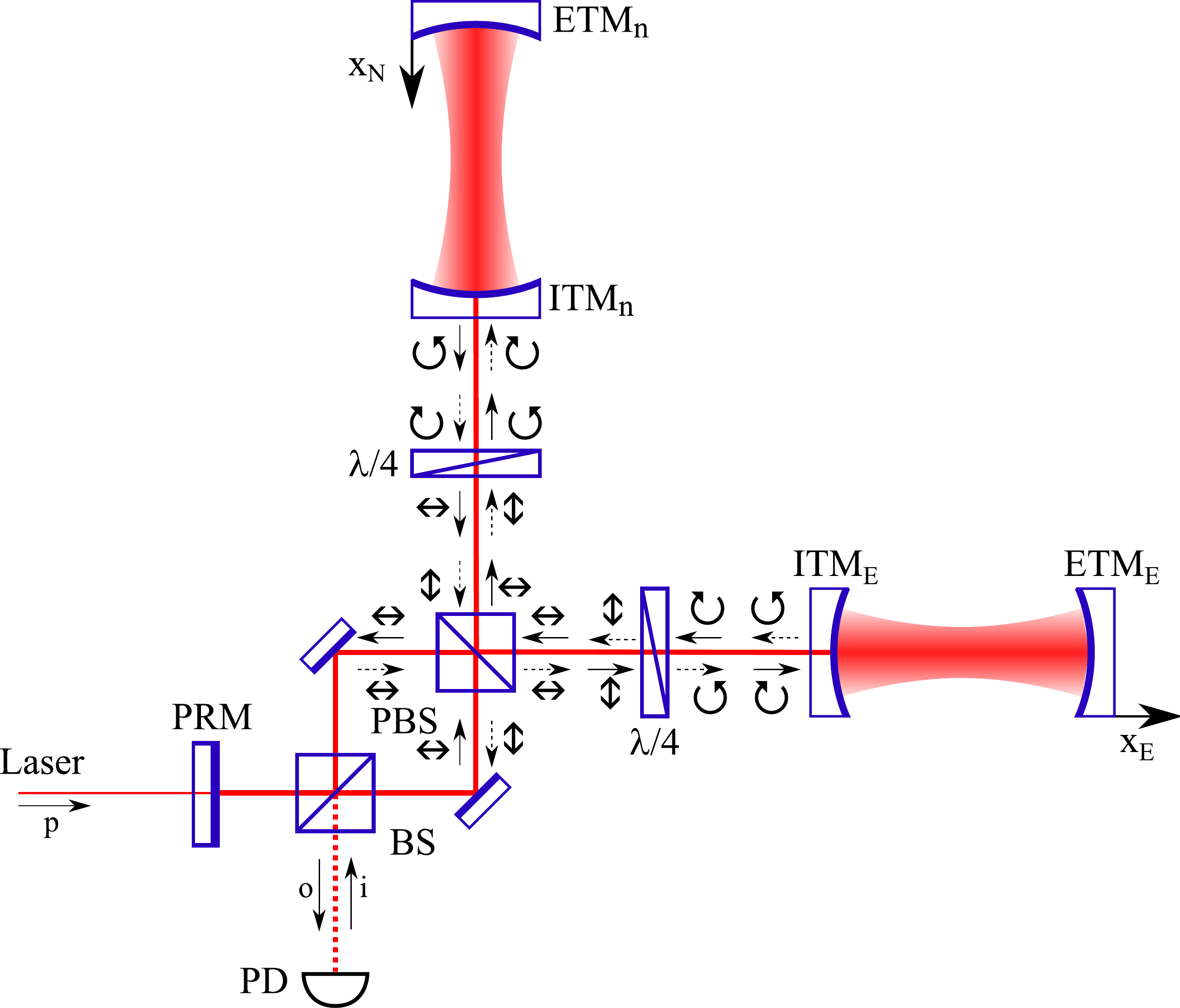}
  \caption{
    Two possible optical realizations of zero area Sagnac speed meter.
    \textit{Left panel:} The ring cavities can be used to spatially separate the ingoing fields from the  outgoing ones, in order to redirect output light from one arm to another one \cite{Chen2002}.
    \textit{Right panel:} The same goal can be achieved idea is realized using optical circulator consisting of the polarization beamsplitter (PBS) and two $\lambda/4$-plates \cite{02a2Kh,04a1Da}.
  }\label{fig:Sagnacfig}
\end{figure*}

Later on, the idea to use the zero area Sagnac interferometer \cite{Byer1996,Byer1999,Byer2000} as a speed meter was proposed independently by Chen and Khalili \cite{Chen2002,02a2Kh}. Further analysis with account for optical losses was performed in Ref.~\cite{04a1Da} and with detuned signal-recycling in Ref.~\cite{MSSDC2005}. Suggested configurations are pictured in Fig.\,\ref{fig:Sagnacfig}. The core idea is that light from the laser gets splitted by the beamsplitter (BS) and directed to Fabry-P\'erot cavities in the arms, exactly as in conventional Fabry-P\'erot---Michelson GW interferometers. However, after it leaves the cavity, it does not go back to the beamsplitter, but rather enters the cavity in the other arm, and only afterwards returns to the beamsplitter, and finally to the photo detector at the dark port. The scheme of Ref.~\cite{Chen2002} uses ring Fabry-P\'erot cavities in the arms to spatially separate ingoing and outgoing light beams to redirect the light leaving the first arm to the second one evading the output beamsplitter. The variant analyzed in Ref.~\cite{02a2Kh,04a1Da} uses polarized optics for the same purposes: light beams after ordinary beamsplitter, having linear ({\it e.g.}, vertical) polarization, pass through the polarized beamsplitter (PBS), then meet the $\lambda/4$ plates that transforms their linear polarization into circular one, and then enter the Fabry-P\'erot  cavity. After reflection from the Fabry-P\'erot  cavity, light passes through $\lambda/4$-plate again, changing its polarization again to linear, but orthogonal to the initial one. As a result, the PBS reflects it and redirects to another arm of the interferometer where it passes through the same stages, restores finally the initial polarization and comes out of the interferometer. With the exception of the implementation method for this round-robin pass of the light through the interferometer, both schemes has the same performance, and the same appellation ``Sagnac speed meter" will be used for them below.

Visiting consequently both arms, the counter propagating light
beams acquire a phase shifts proportional to the sum of the end
mirrors' displacement in both cavities taken with time delay equal
to average single cavity storage time $\tau_{\rm arm}$:
\begin{equation}
    \delta\phi_R \propto x_N(t)+x_E(t+\tau_{\rm arm})\,,\quad
    \delta\phi_L \propto x_E(t)+x_N(t+\tau_{\rm arm})\,.
\end{equation}
After recombining at the beamsplitter and photo detection, the
output signal will be proportional to the phase difference of
clockwise (R) and counter clockwise (L) propagating light beams:
\begin{equation}\label{phi_speedmeter}
 \delta\phi_R - \delta\phi_L \propto [x_N(t)-x_N(t+\tau_{\rm arm})]
   -[x_E(t)-x_E(t+\tau_{\rm arm})]
  \propto \dot x_N(t) - \dot x_E(t) + O(\tau_{\rm arm})
\end{equation}
that, for frequencies $\ll\tau_{\rm arm}^{-1}$, is proportional to
the relative velocity of the interferometer end test masses. The
presence of two counter propagating light beams exciting both arm
cavities in a symmetric way allows to mostly keep the traditional
optical layout of GW interferometers with two large scale high
finesse Fabry-P\'erot  cavities, minimizing the cost and the
operation complexity of this scheme and allowing to switch between
the speed meter and ordinary position meter operation modes.

\subsection{Speed meter sensitivity}

As shown by Chen~\cite{Chen2002}, a mapping exists between the
Sagnac and the Michelson speed meter topology, if tuned signal
recycling is considered: both are characterized by two optical
resonators, with resonant frequencies located symmetrically around
the carrier frequency.  Henceforth, we will use Sagnac speed
meters as representative for different possible speed meter
realizations.

The quantum noise spectral density of a lossless speed meter is similar in form to Eq.\,(\ref{S_PM})
\begin{equation}\label{S_SM}
  S^h_{\rm SM} = \frac{h^2_{\rm SQL}}{2}\,
    \frac{e^{-2r} + [\cot\zeta-{\cal K}_{\rm SM}(\Omega)]^2e^{2r}}{{\cal K}_{\rm SM}(\Omega)} \,,
\end{equation}
where ${\cal K}_{\rm SM}$ is the optomechanical coupling of the
speed meter as given in Sec.\,IID of Ref.\,\cite{Chen2002}
\begin{equation}\label{KSM}
  {\cal K}_{\rm SM}(\Omega) = \frac{4I_c}{I_c^{\rm SQL}}\,\frac{\gamma^4}{(\gamma^2+\Omega^2)^2}\,,
\end{equation}
and
\begin{equation}
  I_c = {8\,I_{\rm BS}}/{T} \,.
\end{equation}
Note that for a given $I_{\rm BS}$, this $I_c$ is twice  that of the position
meter [Cf.~Eq.\,(\ref{I_c_PM})], because here each beam visits both arms sequentially after leaving the beamsplitter.

\begin{figure}[t]
  \includegraphics[width=0.5\textwidth]{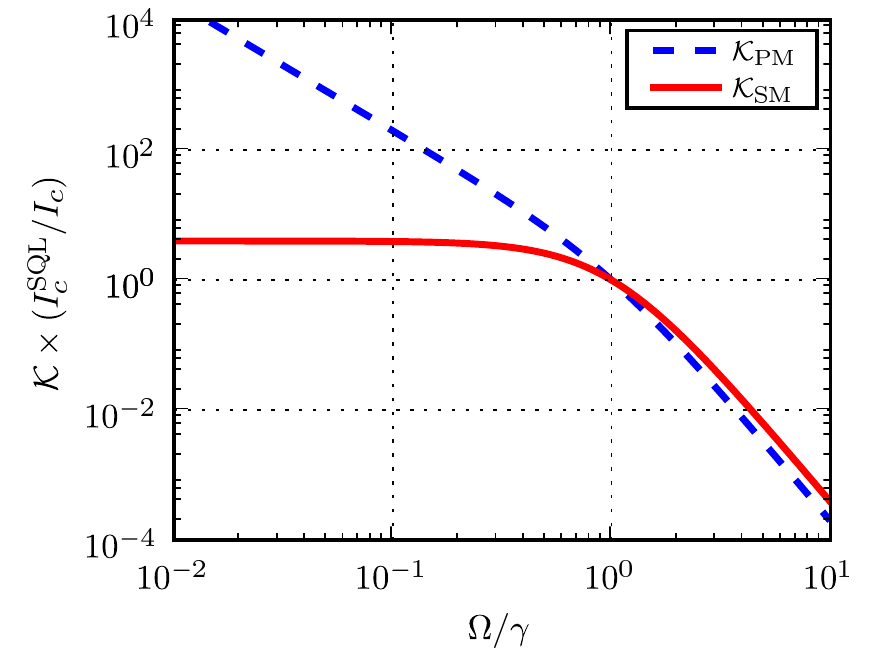}\hfill
  \includegraphics[width=0.5\textwidth]{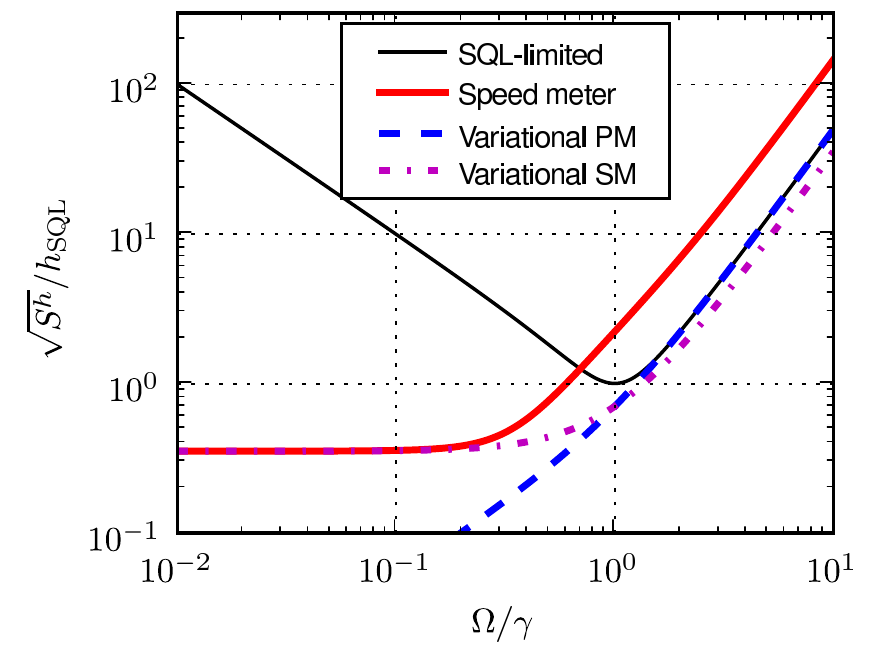}
  \caption{Variational position meter compared with speed meter. Left panel: optomechanical coupling factors for position meter ($\mathcal{K}_{\rm PM}$ and speed meter ($\mathcal{K}_{\rm SM}$) as functions of signal frequency $\Omega$. Right panel: SQL beating factors $\sqrt{S_h}/h_{\rm SQL}$ for low frequency optimized  speed meter interferometer (Speed meter), variational position meter (Variational PM), variational  speed meter (Variational SM), and SQL-limited interferometer. In all cases, $I_ce^{2r}=I_c^{\rm SQL}$.}\label{fig:noloss}
\end{figure}

The key advantage of speed meters over position meters is that at
low  frequencies, $\Omega<\gamma$, $\mathcal{K}_{\rm SM}$ is
approximately constant and reaches a maximum there
\begin{equation}
  {\cal K}_{\rm SM}(\Omega\ll \gamma)\approx {4I_c}/{I_c^{\rm SQL}} \,.
\end{equation}
As a consequence, a {\it
frequency-independent} readout quadrature optimized for low frequencies can be used:
\begin{equation}
  \zeta = \arccot\mathcal{K}_{\rm SM}(0) = \arccot({4I_c}/{I_c^{\rm SQL}}) \,,
\end{equation}
which gives the following spectral density
\begin{equation}\label{S_SM_LF}
  S^h_{\rm SM\,LF} = \frac{h^2_{\rm SQL}(\Omega)}{2 }\left[
    \frac{e^{-2r}}{\mathcal{K}_{\rm SM}(\Omega)}
    + \frac{\Omega^4(2\gamma^2+\Omega^2)^2}{\gamma^8}\mathcal{K}_{\rm SM}(\Omega)e^{2r}
  \right] .
\end{equation}
Here the radiation-pressure noise (second term in bracket) is significantly suppressed in low frequencies ($\Omega \stackrel{<}{_\sim}\gamma$), and $  S^h_{\rm SM\,LF}$  can beat the SQL in a broad frequency band.  By contrast, this is not possible for position meters, which have $\mathcal{K}_{\rm PM} \sim 1/\Omega^2$, see Fig.\,\ref{fig:noloss} (left panel).

Variational readout technique~\cite{02a1KiLeMaThVy} discussed in Sec.\,\ref{ssec:variational} can also be used to enhance the speed meters performance \cite{Purdue2002}. The optimal frequency-dependent readout quadrature is given by
\begin{equation}
  \zeta = \arccot {\cal K}_{\rm SM}(\Omega)\,,
\end{equation}
which gives a noise spectral density without the back action term:
\begin{equation}
  S^h_{\rm SM\,VM} = \frac{h^2_{\rm SQL}e^{-2r}}{2{\cal K}_{\rm SM}(\Omega)} \,.
    \label{S_SM_VM}
\end{equation}

In right panel of Fig.\,\ref{fig:noloss} we plot the SQL-beating factors $\sqrt{S^h}/h_{\rm SQL}$ for low-frequency optimized speed meter (\ref{S_SM_LF}), variational position (\ref{S_PM_VM}) and speed meters (\ref{S_SM_VM}). For comparison, $\sqrt{S^h}/h_{\rm SQL}$ for the SQL-limited interferometer (\ref{S_SQL}) is also given.
One might conclude from these plots that the variational
position meter is clearly better than the speed meter (except for
high frequency area where the most sophisticated scheme of
variational speed meter has some marginal advantage) --- however in the next subsection we demonstrate that optical losses change this picture radically.

\subsection{Optical losses in speed meters}

In speed meters, optical losses in the arm cavities significantly affect the sum noise at low frequencies, even if $1-\eta_{\rm cav}\ll1-\eta$, because the radiation pressure noise component created by the arm cavities losses has frequency dependence similar to the one for position meters (remind that $\mathcal{K}_{\rm PM}/\mathcal{K}_{\rm SM}\to\infty$ if $\Omega\to0$, see Fig.\,\ref{fig:noloss}, left). In this paper, we will use the following expression for the lossy speed meter sum noise (which is still simplified, but takes the losses frequency dependence into account; more detailed treatment of the lossy speed meter can be found in papers \cite{Purdue2002,04a1Da}):
\begin{equation}\label{S_SM_loss}
  S^h_{\rm SM} = \frac{h^2_{\rm SQL}}{2}\biggl\{
    \frac{
      e^{-2r} + \xi_{\rm loss}^4/\sin^2\zeta
      + [\cot\zeta - \mathcal{K}_{\rm SM}(\Omega)]^2e^{2r}
    }{\mathcal{K}_{\rm SM}(\Omega)}
    + \xi_{\rm cav}^4\mathcal{K}_{\rm PM}(\Omega)
  \biggr\} ,
\end{equation}
where $\xi_{\rm cav} = \sqrt[4]{(1-\eta_{\rm cav})/\eta_{\rm cav}}$\,.

The low-frequency optimized detection angle, in presence of loss, is
\begin{equation}
  \zeta = \arccot\frac{\mathcal{K}(0)}{1 + \xi_{\rm loss}^4e^{-2r}}
  = \arccot\frac{4I_c/I_c^{\rm SQL}}{1 + \xi_{\rm loss}^4e^{-2r}} \,,
\end{equation}
which gives
\begin{multline}\label{S_SM_LF_loss}
  S^h_{\rm SM\,LF} = \frac{h^2_{\rm SQL}}{2}\biggl\{
    \frac{e^{-2r} + \xi_{\rm loss}^4}{{\cal K}_{\rm SM}(\Omega)}
    + \frac{{\cal K}_{\rm SM}(\Omega)}{1 + \xi_{\rm loss}^4e^{-2r}}
        \left[\frac{\Omega^4(\Omega^2+2\gamma^2)^2}{\gamma^8}\,e^{2r} + \xi_{\rm loss}^4\right] \\
    + \xi_{\rm cav}^4\mathcal{K}_{\rm PM}(\Omega)
  \biggr\} ,
\end{multline}
[compare with Eq.\,(\ref{S_SM_LF}) and note the additional residual back action term similar to one in Eq.\,(\ref{S_PM_VM_loss})].


For a lossy variational speedmeter, repeating calculations of Sec.\,\ref{ssec:losses}, we obtain
\begin{equation}
  \zeta(\Omega) = \arccot\frac{\mathcal{K}_{\rm SM}(\Omega)}{1 + \xi_{\rm loss}^4e^{-2r}} \,,
\end{equation}
which results in a total noise spectrum of
\begin{equation}\label{S_SM_VM_loss}
  S^h_{\rm SM\,VM} = \frac{h^2_{\rm SQL}}{2}\left[
    \frac{e^{-2r} + \xi_{\rm loss}^4}{{\cal K}_{\rm SM}(\Omega)}
    + \frac{\xi_{\rm loss}^4{\cal K}_{\rm SM}(\Omega)}{1 + \xi_{\rm loss}^4e^{-2r}}
    + \xi_{\rm cav}^4\mathcal{K}_{\rm PM}(\Omega)
  \right] .
\end{equation}
{\rm }It is easy to see that for both, the low-frequency
optimized speed meter and the variational speed meter, optical
losses impose the same fundamental limitation (\ref{xi_loss}) on
the SQL beating factor $\sqrt{S_h}/h_{\rm SQL}$.

\begin{figure*}[t]
  \includegraphics[width=0.50\textwidth]{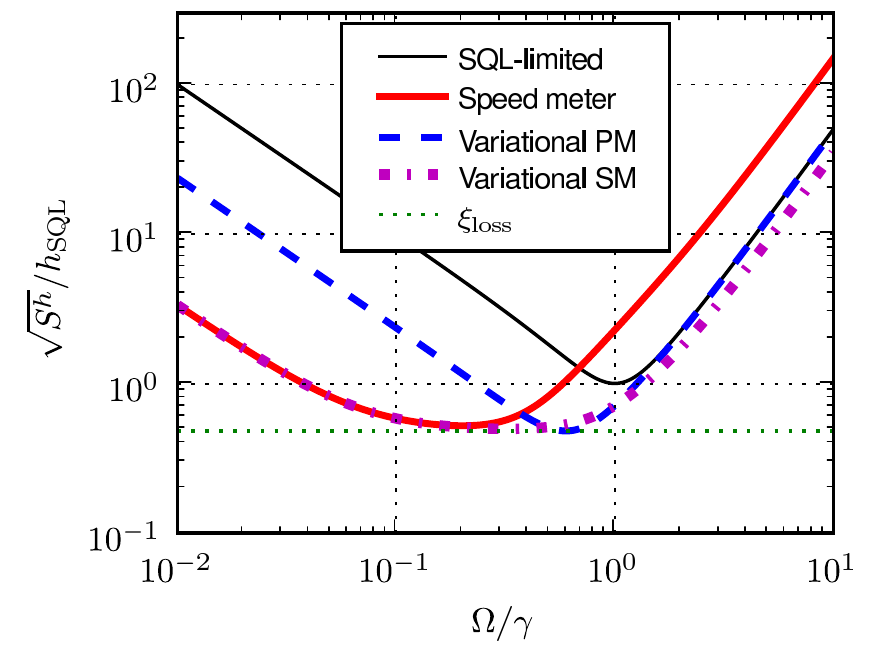}\hfill\includegraphics[width=0.50\textwidth]{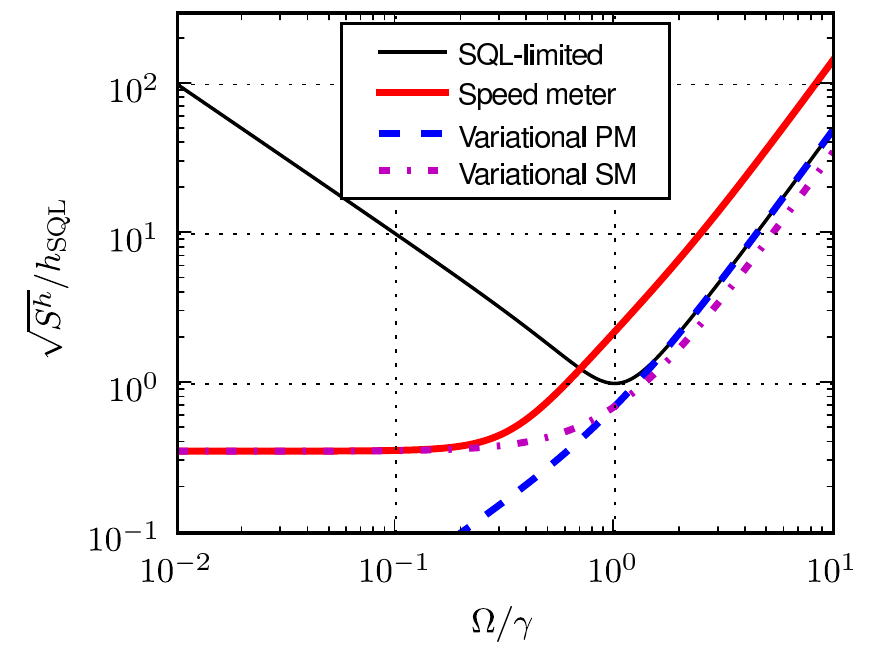}
  \caption{
    SQL beating factors $\sqrt{S_h}/h_{\rm SQL}$, in presence of optical losses ($\eta=0.95$, $\eta_{\rm cav}=1-10^{-3}$), for: low frequency optimized Sagnac speed meter interferometer (Speed meter), variational position meter (Variational PM), variational Sagnac speed meter (Variational SM). Left panel:  $I_c=I_c^{\rm SQL}$ and $r=0$ (no squeezing). Right panel: $I_ce^{2r}=0.1I_c^{\rm SQL}$ and $e^{2r}=10$. On both panels, $\sqrt{S_h}/h_{\rm SQL}$ of lossless SQL-limited interferometer with $I_ce^{2r}=I_c^{\rm SQL}$ is plotted for comparison (SQL-limited).
  }\label{fig:loss}
\end{figure*}

In Fig.\,\ref{fig:loss}, the factors $\sqrt{S_h}/h_{\rm SQL}$ for low frequency optimized lossy speed meter (\ref{S_SM_LF_loss}) and lossy variational position and speed meters (\ref{S_PM_VM_loss}, \ref{S_SM_VM_loss}) are plotted, for $I_c=I_c^{\rm SQL}$ without squeezing (left panel) and $I_c=0.1I_c^{\rm SQL}$ with 10 dB squeezing (right panel). In both cases, realistic parameter $\eta=0.95$ and $\eta_{\rm cav}=1-10^{-3}$ are assumed. These plots indicate that in presence of optical loss, the speed meter provides better broadband low-frequency sensitivity than variational position meters.

These plots also suggest that the
interferometer's bandwidth $\gamma$ should be larger than the band of interest.  This sets a minimum requirement for  $I_c^{\rm SQL}$ [Cf.~Eq.~(\ref{I_c_SQL})], and thus the necessary optical circulating
power $I_c$. However, if a speed meter is only used for lower frequencies, the necessary  optical power may not be very high, even without the squeezing. For
example, $M=200\,{\rm kg}$, $L=10\,{\rm km}$, and
$\gamma=100\,{\rm s}^{-1}$, then $I_c^{\rm SQL}\approx80\,{\rm kW}$
--- only four times as high as in the initial LIGO detector.
On the other hand, for higher working frequnecies, the power requirements increase sharply. For example, $\gamma=2\pi\times100\,{\rm s}^{-1}$ requires  $I_c^{\rm SQL}\approx20\,{\rm MW}$.

\subsection{Detuned speed meter and optical inertia}\label{ssec:detuning}

So far we considered only resonance tuned Sagnac speed meter.  A convenient way of detuning the  Sagnac interferometer is to use the signal-recycled topology originally proposed by Meers \cite{Meers1988}: an additional mirror is placed at the dark output port of the interferometer, reflecting parts of the signal modulation fields back into the interferometer and forming a cavity together with the input mirrors of the interferometer's arm cavities. When this new cavity is neither resonant nor anti-resonant with respect to the carrier frequency, the signal mode of the interferometer also becomes detuned, and the interferometer has a peak sensitivity around a non-zero  resonant frequency. Nearly all second generation GW detectors will make use of this technique.

In a detuned position meter, the optomechanical coupling induces a restoring force onto the differential mode of the arm-cavities mirrors \cite{Buonanno2001,Buonanno2002,Buonanno2003}: motion of mirrors produce phase modulation of the out-going light, which, through detuning, gets converted into amplitude modulation and acts back onto the mirrors as a position-dependent force.  This {\it optical spring} can shift the pendulum frequency of the mirror and move it into the detection band.  In this case the interferometer's sensitivity has two peaks, one around the original optical resonance due to detuning, the other due to the upshifted mechanical resonance.
A full explicit expression for the quantum noise spectral density
can be found in Eq.~(37) of Ref.~\cite{Buonanno2003}.

The detuned Sagnac interferometer has also two resonances but both
are of optical nature, due to the existence of two optical
resonators in the system; differently from position meters, the
optomechanical coupling in a Sagnac modifies the dynamical mass of
the mirrors, adding an {\it optical inertia}~ \cite{MSSDC2005}. In
a Sagnac, amplitude modulation $a$ entering from the dark port act
successively on both cavities, and in low frequencies
\begin{equation}
  F_{\mathrm{RP}}^{\mathrm{N}} - F_{\mathrm{RP}}^{\mathrm{E}}
  \propto \dot a (t)\,.
\end{equation}
In a detuned situation, $a$ contains some of the out-going phase modulation, which is proportional to $\dot{x}_{\rm N} - \dot{x}_{\rm E}$, and therefore the radiation-pressure force contains  the mirrors' {\it acceleration:}.
\begin{equation}
  F_{\mathrm{RP}}^{\mathrm{N}} - F_{\mathrm{RP}}^{\mathrm{E}}
    \propto -M_{\rm opt}\left(\ddot x_{\mathrm{N}} - \ddot x_{\mathrm{E}}\right) .
\end{equation}
Here $M_{\rm opt}$ is the effective optical inertia which can be either positive or negative.
Because this treatment is true for all frequencies below the optical resonances, it is possible to achieve $M+M_{\rm opt} \ll M$ for a broad frequency band --- this leads to a broadband amplification of GW signal, and may allow us to circumvent the fundamental limit of \eqref{xi_loss} imposed by optical losses.

\section{Intracavity measurement}\label{sec:intracavity}

\subsection{The idea of ``optical bars'' and ``optical lever'' topologies}

QND versions of traditional interferometer schemes, like, for example, variational meter and quantum speed meter, considered in the previous sections, allow to subtract the back-action noise from the output signal. As a result, only measurement noise remains (in lossless case) in the output signal which has spectral density with the general structure demonstrated by Eqs.\,(\ref{S_PM_VM}, \ref{S_SM_VM}).

Consider dependence of this spectral density on the arms length $L$. On he one hand, it is directly proportional to $L$ through the opto-mechanical coupling factor $\mathcal{K}$, see Eqs.\,(\ref{KPM}, \ref{I_c_PM}, \ref{KSM}). On the other hand, it is inversly proportional to $L$ through the factor $h_{\rm SQL}^2$, see Eq.\,\eqref{eq:SQL}. The physical meanings of these dependencies are evident: the shorter are the arms, the better is {\it displacement} sensitivity, for given values of the interferometer bandwidth $\gamma$ and the circulating power $I_c$. But signal displacement itself is proportional to the length of the arms:
\begin{equation}\label{h2x}  x = Lh/2 \,, \end{equation}
and that eventually leads to decrease of overall sensitivity to {\it GW signal} $h$ with the arm length decrease.

\begin{figure*}[t]
\includegraphics[width=0.42\textwidth]{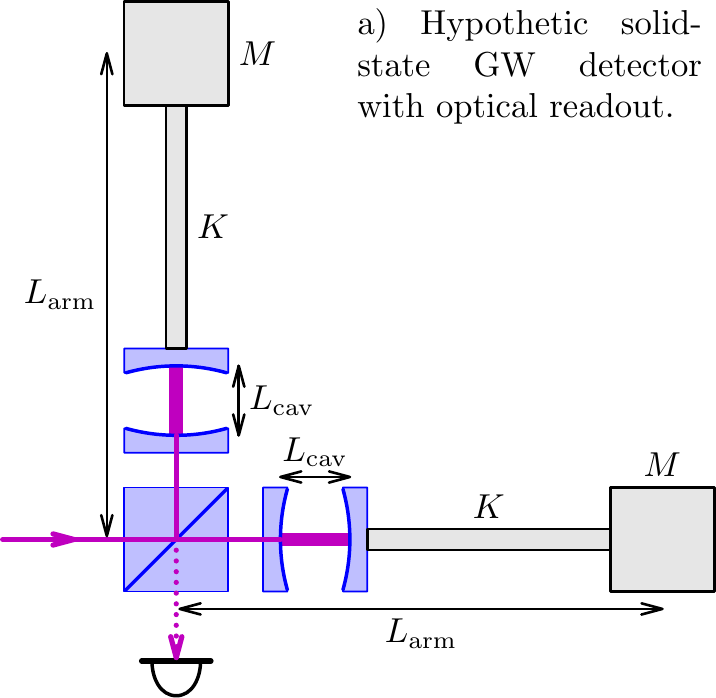}\qquad
\includegraphics[width=0.42\textwidth]{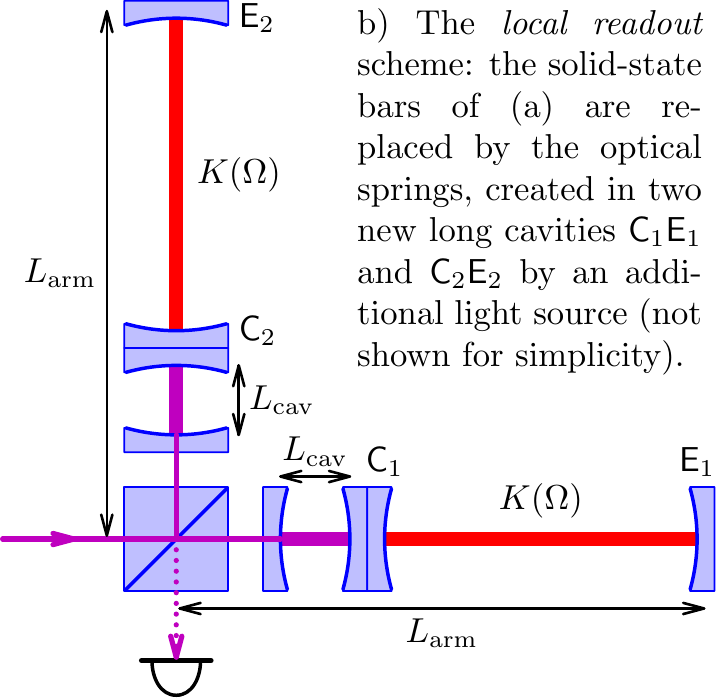}

\medskip
\includegraphics[width=0.42\textwidth]{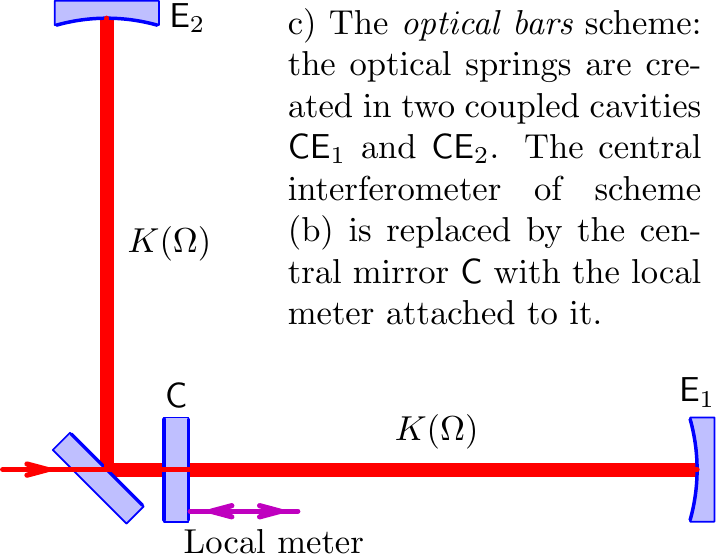}\qquad
\includegraphics[width=0.42\textwidth]{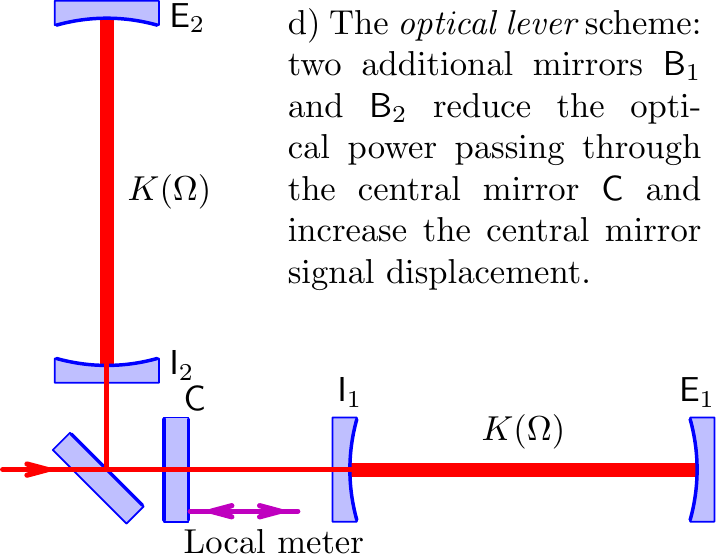}
\caption{Transition from the solid-state bar detector to the optical bars/optical lever intracavity readout scheme.}\label{fig:bars}
\end{figure*}

This reasoning implies silently that the length $L$ of Eq.\,(\ref{h2x}) and the one met in $S^x$, are the same lengths equal to each other. This is the case for contemporary laser interferometric GW detectors, but is absolutely not the case \textit{e.g.} for solid-state detectors (see review article \cite{00a1JuBlZh} and references therein for details on solid-state GW detectors). Consider now a hypothetic bar detector with optical readout shown in Fig.\,\ref{fig:bars}(a). Here two stiff bars transfer the GW induced displacement of the test masses to the end mirrors of the standard Fabry-P\'erot--Michelson interferometer. In principle, in this case the arms length $L_{\rm arm}$ that matters for GW strain to test mass displacement transformation [the one in Eq.\,(\ref{h2x})], can be much longer than the cavities length $L_{\rm cav}$ (which appears in $S^x$), and thus the circulating optical power in the interferometer can be reduced by a factor of $L_{\rm cav}/L_{\rm arm}$ without any loss of sensitivity.

The bars in this scheme should be rigid enough to uniformly transfer the GW induced force to the mirrors of FP cavity in the whole frequency range of interest that implies the requirement on the bar rigidity $K$: the lowest eigenfrequency of the bar elastic oscillations has to be higher than the upper expected GW frequency $\Omega$. Of course, for kilometer-scale arms of modern GW detectors, this condition can not be satisfied with ordinary solid-state bars. However,  one can use two long Fabry-P\'erot cavities instead where an optical spring can be created by a means of auxiliary detuned laser pumping \cite{99a1BrKh,01a2Kh,Buonanno2002}, see Fig.\,\ref{fig:bars}(b). In essence, this is the so-called {\it local readout} topology considered in \cite{Rehbein2007}.

The first proposed intracavity readout scheme, {\it optical bars} \cite{97a1BrGoKh}, was based on the same principle. Fig.\,\ref{fig:bars}(c) demonstrates how it works. Here optical springs are created in the system of two cavities ${\sf CE}_1$ and ${\sf CE}_2$ coupled by means of partly transparent central mirror {\sf C}. Optical eigenfrequencies of this system form a set of doublets, with frequencies in each doublet separated by the sloshing frequency
\begin{equation}
  \Omega_B = \frac{c\sqrt{T_{\sf C}}}{L}  \,,
\end{equation}
where $T_{\sf C}$ is the central mirror power transmissivity. If the upper frequency mode of some of the doublets is pumped then optical field acts similarly to two springs where one is created between the mirrors ${\sf E}_1$ and {\sf C} while the second one (L-shaped) appears between the mirrors ${\sf E}_2$ and {\sf C}. The corresponding rigidities are given by expressions
\begin{equation}\label{K_optbar}
  K(\Omega) \approx \frac{4\omega_oI_c^{\rm bar}}{Lc}\,\frac{\Omega_B}{\Omega_B^2-\Omega^2} \,,
\end{equation}
where $I_c^{\rm bar}$ is the optical power circulating in each of these cavities. Similar to previous two schemes, these springs transform the GW induced displacement of the end mirrors ${\sf E}_1$, ${\sf E}_2$ directly into the displacement of central mirror ${\sf C}_1$. This displacement, in turn, has to be monitored by a means of some {\it local meter} attached to this mirror. It is this local meter that plays the same role as the small Fabry-Perot--Michelson interferometer of Figs.\,\ref{fig:bars}(a,b).

Dynamics of the {\it optical bars} system with its three movable mirrors coupled to each other by means of frequency-dependent rigidity (\ref{K_optbar}), is quite sophisticated. Depending on the values of mirrors masses, optical power $I_{\sf C}^{\rm bar}$ and sloshing frequency $\Omega_B$, there could be either two resonances (mechanical and  optical \cite{Buonanno2002}), or one second-order resonance \cite{01a2Kh}, which can be used to significantly increase the central mirror signal displacement in narrow band. However, in broadband regimes with flat mechanical transfer function, which are most interesting for GW experiments, the signal displacement of the central mirror meter is the same as those of the end mirrors:
\begin{equation}\label{x_s_optbar}
  x_s(\Omega) = \frac{M_{\sf E}}{2M_{\sf E}+M_{\sf C}}\,Lh(\Omega) \,,
\end{equation}
where $h(\Omega)$ is the GW strain signal (see the next subsection for more detail).

In the article \cite{02a1Kh}, an improved version of the {\it optical bars} scheme, which allows to increase $x_s$, was proposed. It differs from the original one by two additional mirrors ${\sf I}_1$ and ${\sf I}_2$, see Fig.\,\ref{fig:bars}(d), which turn the arms into two Fabry-P\'erot cavities. This scheme was called {\it optical lever} because the signal displacement gain \rem{which} it provides is similar to the one which can be obtained by ordinary mechanical lever with arms lengths ratio equal to
\begin{equation}
  \digamma = \frac{2}{\pi}\,\mathcal{F} \,,
\end{equation}
where $\mathcal{F}$ is the finesse of these Fabry-P\'erot cavities. With this addition, Eq.\,(\ref{x_s_optbar}) takes the following form:
\begin{equation}\label{x_s_optlever}
  x_s(\Omega) = \frac{\digamma M_{\rm arm}}{2M_{\rm arm} + \digamma^2M_{\sf C}}\,Lh(\Omega) \,,
\end{equation}
where
\begin{equation}
  M_{\rm arm} = \frac{M_{\sf E}M_{\sf I}}{M_{\sf E} + M_{\sf I}}
\end{equation}
is the reduced mass of the new arm cavities mirrors. 

\subsection{The optical bars regimes}

The optical bars principle, as one can see from the above subsection, allows to reduce the task of monitoring the relative displacement of two test masses separated by kilometer-scale distance to a much easier task of detecting the same, or even amplified (in case of ``optical lever'' scheme) motion of the local mirror {\sf C} with respect to some reference object that has no optical coupling to the intracavity field.

The evident price is the necessity to have additional long cavities intended for provision of optical springs. The optical power requirements for these cavities were estimated in Sec.\,4 of paper \cite{98a1BrGoKh}. Assuming that the mass of the central mirror is small enough to maximize its mechanical response and thus the measured displacement:
\begin{equation}
  M_{\sf C} < \frac{M_{\rm arm}}{\digamma^2} \,,
\end{equation}
and that GW signal frequencies $\Omega$ are below the sloshing frequency: $\Omega<\Omega_B$, one can distinguish three working regimes that are defined by available pumping power.

\paragraph{Strong pumping:}

\begin{equation}\label{K_strong}
  K(0) = \frac{4\omega_oI_c^{\rm bar}}{Lc\Omega_B} > M_{\rm  arm}\Omega^2 \,.
\end{equation}
In this regime, optical bars work as absolutely stiff rods, transfering 100\% of signal displacement of the end mirror to the central one [see Eqs.\,(\ref{x_s_optbar}, \ref{x_s_optlever})]. The opposite is also true, namely the local meter back action force on the central mirror perfectly passes to the end mirrors. The simplest option for the local meter in this case is just an ordinary position meter with sensitivity correponding to the SQL for the equivalent mass $M_{\rm arm}/\digamma^2$:
\begin{equation}
  S_x^{\rm local}(\Omega) \simeq \frac{2\hbar\digamma^2}{M_{\rm arm}\Omega^2} \,.
\end{equation}
Due to amplification factor $\digamma$, the net sensitivity in this case corresponds to the SQL for the mass $M_{\rm arm}$:
\begin{equation}\label{S_x_SQL}
  S_x(\Omega) \simeq \frac{2\hbar}{M_{\rm arm}\Omega^2} \,.
\end{equation}
Of course, using a QND local meter with sub-SQL sum noise it is possible to get corresponding improvement of the overall sensitivity.

The optical power required to create such stiff optical bars is relatively high: it follows from Eq.\,(\ref{K_strong}) that
\begin{equation}
  I_c^{\rm bar} > \frac{M_{\rm arm}Lc\Omega_B^3}{4\omega_o}.
\end{equation}
Note that the right hand side of the above relation is equal to the circulating power $I_c^{\rm SQL}$ needed to reach the SQL in a conventional interferometer with the same arms length, mirrors mass, and with bandwidth $\gamma$ equal to $\Omega_B$ [see Eq.\,(\ref{I_c_SQL})]. It is important, however, that $I_c^{\rm bar}$ does not depend on the desired sensitivity (which is defined by the local meter), while in the ordinary topologies, the power required increases as square of the signal displacement which has to be detected, see, {\it e.g.}, Eqs.\,(\ref{S_PM_VM}, \ref{S_SM_VM}).

\paragraph{Intermediate pumping:}
\begin{equation}\label{K_inter}
  \digamma^2M_{\sf C}\Omega^2 < K(0) = \frac{4\omega_oI_c^{\rm bar}}{Lc\Omega_B}
    < M_{\rm arm}\Omega^2 \,.
\end{equation}
This regime, being the most interesting one, was considered in detail in paper \cite{03a1Kh}. Therein the stiffness of optical bars suffices to satisfy Eqs.\,(\ref{x_s_optbar}, \ref{x_s_optlever}), but is not enough to transfer the local meter back action force to the heavy end mirrors. As a result, the local meter ``sees'' an additional back action noise. However, this noise can be excluded from the local meter output signal using frequency-independent cross correlation of the measurement and back-action noises. This kind of noise correlation can be created relatively easy using homodyne detection of the local meter output with properly set homodyne angle $\zeta$. In this case, the same sensitivity (\ref{S_x_SQL}) as in the previous case can be obtained, but with significantly less optical power circulating in the ``optical bar'' cavities. The use of QND local meter with sub-SQL sum noise will be able to provide proportionally better sensitivity.

It was shown in paper \cite{03a1Kh}, that the most fundamental limitation on optical power in this regime is imposed by the power fluctuations in the ``optical bar'' cavities, created by optical losses. The measure of perturbation of central mirror displacement due to radiation pressure fluctuations associated with this loss source can be provided by the following spectral density:
\begin{equation}\label{S_OB_loss}
  S_{\rm loss} \simeq \frac{\hbar Lc}{4\omega_oI_c^{\rm bar}}\,\gamma \,.
\end{equation}
This expression has a structure similar to the displacement noise spectral density $S^x$ of the ``ordinary'' interferometer with $\gamma$ replaced by some frequency dependent factor of the same order of magnitude. However, in the latter case the condition $\gamma\gtrsim\Omega_{\rm GW}$ has to be fulfilled where $\Omega_{\rm GW}\sim10\hyphen100\,{\rm s}^{-1}$ is the characteristic GW signal frequency, while in the former case $\gamma$ is limited only by optical losses in the arm cavities:
\begin{equation}
  \gamma = \frac{cA^2}{4L}
\end{equation}
and, for $L\sim10\,{\rm km}$ and $A^2\sim10^{-5}$, it can be as small as $\gamma\sim0.1\,{\rm s}^{-1}$. This bandwidth corresponds to $2\hyphen3$ orders of magnitude smaller noise spectral density for the same optical power or, alternatively, to $2\hyphen3$ orders of magnitude lower optical power for the same sensitivity.

\paragraph{Weak pumping:}
\begin{equation}\label{K_weak}
  K(0) = \frac{4\omega_oI_c^{\rm arm}}{Lc\Omega_B} < \digamma^2M_{\sf C}\Omega^2 \,.
\end{equation}
In this regime, central mirror signal displacent does not reach the level of Eqs.\,(\ref{x_s_optbar}, \ref{x_s_optlever}). It means that more sensitive QND local meter is required. Taking into account the fact that optical power reduction automatically leads to increased noise (\ref{S_OB_loss}), this regime is admittedly of little interest.

\subsection{Implementation issues}

It follows from the above estimates that central mirror {\sf C}, while being relatively small, has to tolerate significant incident optical power. In case of bare {\it optical bars} scheme, the reasonable mirror mass and optical power should be of the order of several kilograms and several hundred kilowatts, respectively. In the {\it optical lever} scheme, light power impinging on the central mirror is $\digamma$ times smaller, but the mass has to be $\digamma^2$ lighter (down to sub-gram range). Fortunately, power of light {\it passing through} the central mirror can be decreased significantly, if the mirror with both front surfaces carrying reflective coatings was used. The Fabry-P\'erot etalon formed by these two coatings should be tuned in anti-resonance. Then the overall amplitude transmissivity of the etalon will be equal to that of single coating transmissivity squared, and the light power inside the etalon will be reduced by the factor equal to its finesse.

\begin{figure}
  \includegraphics[clip,viewport=100 0 500 280,width=0.7\textwidth]{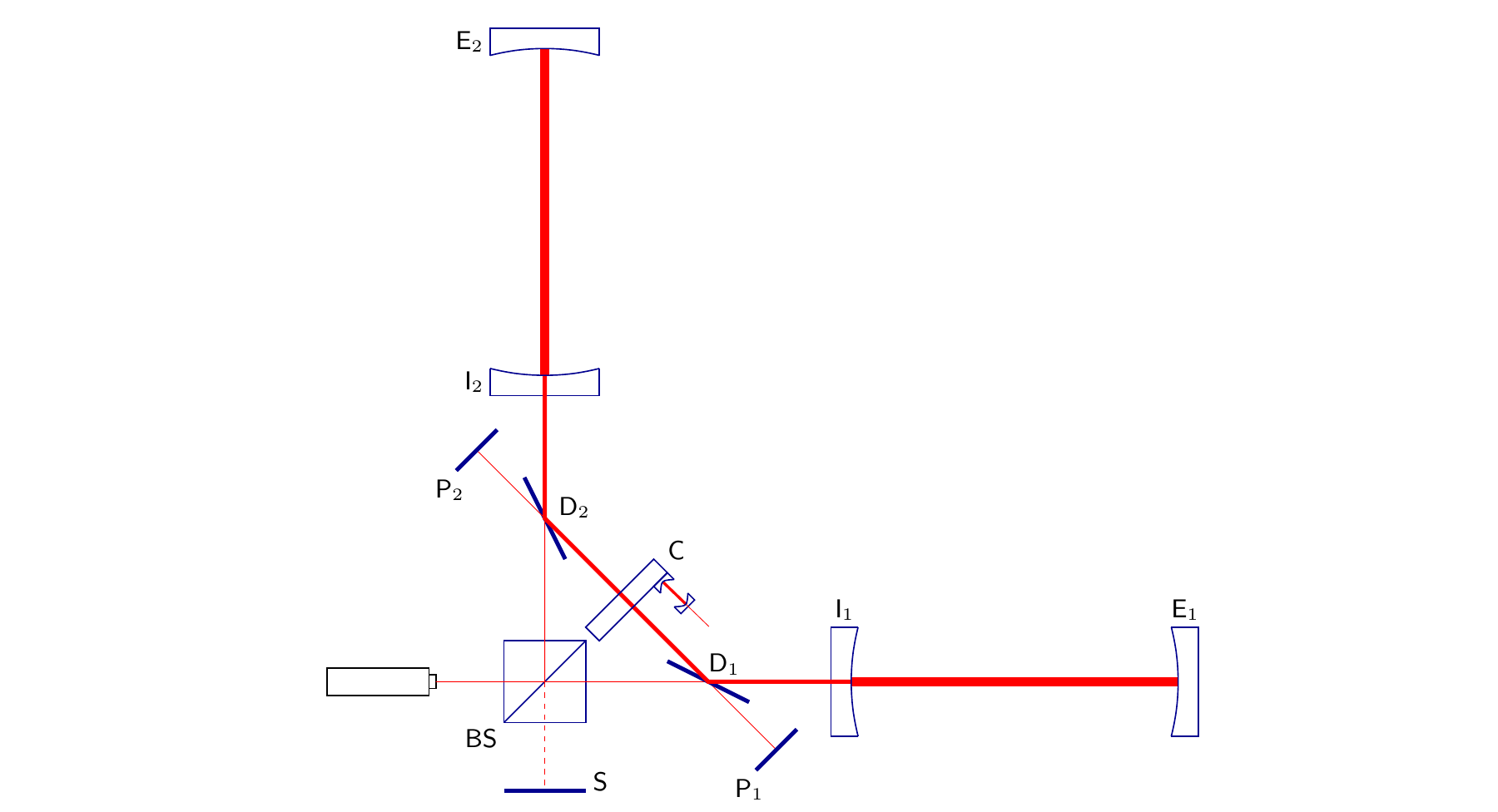}\hfill
  \includegraphics[width=0.3\textwidth]{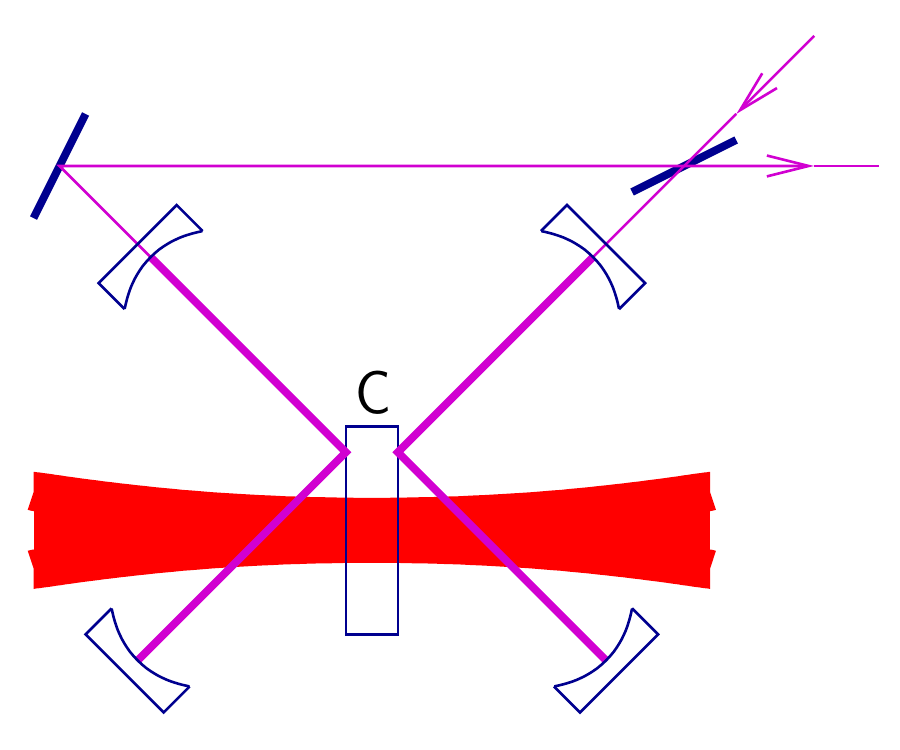}
  \caption{Left: practical version of the {\it optical lever} intracavity scheme. Right a possible topology of the local meter.}\label{fig:optbar2005}
\end{figure}

More implementation issues were addressed in paper \cite{06a1DaKh}, where a practical version of the intracavity detector based on {\it optical lever} principle shown in Fig.\,\ref{fig:optbar2005}, was considered. The following improvements were proposed.

\paragraph{Symmetrization of the topology.}

The evident disadvantage of simple schemes shown in Fig.\,\ref{fig:bars}(c,d) is their asymmetry: pumping power enters  the ``north'' (vertical on the picture) arm first and only then, through the coupling mirror {\sf  C}, the ``east'' one. Due to this asymmetry the input optical field amplitude fluctuations will create differential pondermotive  force acting on the central mirror and mimicing gravitational-wave signal. In order to eliminate this effect, a symmetric power injection scheme has to be used. It consists of the beamsplitter {\sf BS}  and two power injection mirrors ${\sf D}_1$ and ${\sf D}_2$ placed symmetrically on the both sides of the central mirror {\sf C}.

\paragraph{Power recycling mirrors.}

Without power recycling mirrors ${\sf P}_1$, ${\sf P}_2$ one quarter of input power is reflected from the mirrors ${\sf D}_1$ and ${\sf D}_2$ back to the laser, the other quarter is reflected sidewards, and only the remaining half enters the scheme. The mirrors ${\sf P}_1$, ${\sf P}_2$ cancel both reflected beams and increase twice the circulating power inside the scheme twofold, for the same value of input power.

\paragraph{Signal recycling mirror.}

The power injection mirrors ${\sf D}_{1,2}$ finite transmittances create an additional source of optical losses in the scheme. It can be removed using the symmetry of the scheme. Indeed, similar to traditional interferometric gravitational-wave detectors topology, the mean value of optical power circulating inside the scheme depends on the bandwidth of the symmetric optical mode which is coupled to the ``western'' port of the beamsplitter, and the detector sensitivity depends on the bandwidth of anti-symmetric mode which is coupled to ``south'' port of the beamsplitter. The only difference from the traditional topology is in the anti-symmetric mode bandwidth value that has to be as small as possible for intracavity topology in contrast with that of conventional interferometers where it has to be close to GW signal frequency $\Omega$ to provide optimal coupling to the output photodetector. Therefore, high-reflectivity signal recycling mirror {\sf S}  has to be placed in the ``south'' port as shown in Fig.\,\ref{fig:optbar2005}.

\section{Discussion}\label{sec:discussion}

Our estimates show that the use of speed meter topology could
provide at least two to threefold reduction of quantum noise level
in relatively wide frequency band with respect to the level of the
free mass Standard Quantum Limit, with this limitation mostly
owing to optical losses in readout photodetectors. Using injection
of squeezed vacuum into the interferometer dark port, it is
possible to  overcome the SQL by 4-5 times, but in narrower band
of frequencies (see formula (\ref{xi_loss}) and the discussion
thereunder). Optical losses in arm and filter cavities for
variational readout schemes, though being order or two of
magnitude lower than those of photodetectors, show up in low
frequencies due to the inherent frequency dependence they enter
the final quantum noise spectral density, thus limiting the useful
frequency bandwidth of the interferometer.

However, as we have shown, the variational speed meter is less
succeptible to this loss source than the position meter, thus
making the former evident candidate for implenmentation, given the
traditional Michelson-like topology and only slight deviation from
well established and robust techniques of the first generation
detectors are chosen as a strategy of development towards the
third generation devices. Worth noting is the fact, that due to
the flat low-frequency behavior of the speed meter quantum noise
spectral density, even modest SQL-beating factor leads to a huge
sensitivity gain compared to conventional SQL-limited
interferometers, see for example Fig.\,\ref{fig:loss}, where this
gain reaches two orders of magnitude.

An important issue is the optical power that has to circulate in the interferometers to achieve sub-SQL sensitivity. It depends sharply (as $\sim\Omega^3$) on the signal frequency $\Omega$ {\rm (in part, because the free mass SQL for GW signal $h$ (and displacement $x$) itself depends on frequency as $\sim\Omega^{-1}$)} and could potentially be a serious problem for detectors, targeted at higher frequencies. The evident way to relax this requirement is to use squeezing. {\rm However}, this solution is also severely limited by optical losses in photodetectors, deteriorating injected light squeeze factor as a result of mixing with loss related vacuum fields. For example, moderate optimistic value of the unified quantum efficiency $\eta=0.95$ limits the effective squeeze factor to $\lesssim 13\,{\rm dB}$.

Another possibility to increase sensitivity we consider here is to use intracavity readout schemes. The main advantage of these schemes is that their sensitivity can be increased without increase of optical power. Only some fixed threshold value of the power is required in order to create sufficienly stiff optical springs. This value is governed mainly by optical losses and can be much smaller than $I_c^{\rm SQL}$. The gain in the optical power is proportional to the ratio of interferometer bandwidths required for ordinary and intracavity readout schemes respectively. In the first case, the bandwidth has to be of the same order of magnitude as the desired GW signal band, namely $10\hyphen100\,{\rm s}^{-1}$. In the second one, it is limited only by the mirrors absorption and can be as small as $0.1\,{\rm s}^{-1}$ [see discussion around equation (\ref{S_OB_loss})].

An important implementation advantage of the intracavity schemes is that costly part of the setup, namely kilometer-scale optical cavities, is, in essence, nothing more than a simplified version of the ``ordinary'' signal-recycled interferometer without any readout optics, except for those used for the scheme locking. On the other hand, the the intracavity scheme performance is mostly defined by the sensitivity of its second, cheaper because of small size, but completely not developed part: the local meter. It is easy to see, that in the flat sentitivity area $\Omega<\Omega_B$, the SQL beating factor of the whole scheme is equal to one of the local meter. Therefore it has to make use of some QND technique described in our review, or others not mentioned here due to limited scope of this article.

 Another important issue is classical noises in the local meter. On the one hand, smaller test mass means increased suspension and mirror surface noises (due to smaller laser beam radius). On  the other hand, the optical lever topology allows to increase the signal displacement of the mirror. This issue was discussed briefly in the paper \cite{06a2Kh}, where it was shown that reducing the mirror mass at least does not degrade the signal to noise ratio.

It is not obligatory for the local meter to be an optical device, moreover it could be a SQUID-based microwave QND-meter, like speed meters of \cite{90a1BrKh,00a1BrGoKhTh}, or some other high precision superconductive sensor. More traditional for laser gravitational-wave detectors optical sensors with time-domain variational readout (see \cite{06a1DaKh} and references therein) could also be considered as an options.

An interesting intermediate variant which does not differ so radically from traditional schemes is a detuned signal-recycled Michelson interferometer where the differential motion of the end mirrors is coupled to the differential motion of the input ones by the optical springs, and the input mirrors motion is monitored by means of the second pump (the local readout scheme \cite{Rehbein2007}).  If the output port of the main interferometer is closed, then Eqs.\,(\ref{K_optbar}, \ref{x_s_optbar}) also hold with $\Omega_B$ replaced by optical resonance frequency \cite{Buonanno2002}, $I_c^{\rm bar}$ by $I_c$, $M_{\sf C}$ by the mass of the input mirrors and $M_{\sf E}$ by the reduced mass of the end mirrors.

Summing up, we can conclude that intracavity schemes can potentially beat the SQL by larger factor than more traditional extracavity detectors using much smaller circulating optical power, given substantial research effort is put on development of high precision small scale local meter.

\section*{Acknowledgment}

We would like to thank the AEI-Caltech-MIT-MSU MQM discussion group for fruitful discussions and useful comments. Special thanks to Vladimir Braginsky for sharing his invaluable experience and point of view on the way the future gravitational wave detectors should develop. Research of Y.C. is supported by NSF grants  PHY-0653653 and PHY-0601459, as well as the David and Barbara Groce Startup Fund at Caltech.  Research of S.D. is supported by the Alexander von Humboldt Foundation. Research of H.M.-E. is supported by the EC under FP7, Grant Agreement 211743.


\begin{thebibliography}{10}

\bibitem{Abramovici1992}
{A.Abramovici {\it et al}},
\newblock Science {\bf 256}, 325 (1992).

\bibitem{LIGOsite}
\url{http://www.ligo.caltech.edu}.

\bibitem{Ando2001}
{Ando M. {\it et al.}},
\newblock Physical Review Letters {\bf 86}, 3950 (2001).

\bibitem{VIRGOsite}
\url{http://www.virgo.infn.it/}.

\bibitem{Willke2002}
{B.Willke \it{et al.}},
\newblock Classical and Quantum Gravity {\bf 19}, 1377 (2002).

\bibitem{GEOsite}
\url{http://geo600.aei.mpg.de}.

\bibitem{TAMAsite}
\url{http://tamago.mtk.nao.ac.jp}.

\bibitem{Thorne2000}
{K.S.Thorne},
\newblock The scientific case for mature ligo interferometers, 2000,
\newblock {LIGO document P000024-00-R
  (www.ligo.caltech.edu/docs/P/P000024-00.pdf)}.

\bibitem{Fritschel2002}
{P.Fritschel},
\newblock {Second generation instruments for the Laser Interferometer
  Gravitational-wave Observatory (LIGO)},
\newblock in {\em {Gravitational Wave Detection, Proc.~SPIE}}, volume
  {4856-39}, page 282, 2002.

\bibitem{Smith2009}
{J R Smith for the LIGO Scientific Collaboration},
\newblock arXiv:0902.0381  (2009).

\bibitem{Acernese2006-2}
{F.Acernese et al},
\newblock J. Phys.: Conf. Ser. {\bf 32}, s223 (2006).

\bibitem{Willke2006}
{B.Willke et al},
\newblock Classical and Quantum Gravity {\bf 23}, S207 (2006).

\bibitem{LCGTsite}
\url{http://www.icrr.u-tokyo.ac.jp/gr/LCGT.html}.

\bibitem{92BookBrKh}
{V.B.Braginsky, F.Ya.Khalili},
\newblock {\em Quantum Measurement},
\newblock Cambridge University Press, 1992.

\bibitem{Freise2009}
{A. Freise, S. Chelkowski, S. Hild, W. Del Pozzo, A. Perreca and A. Vecchio},
\newblock Classical and Quantum Gravity {\bf 26}, 085012 (2009).

\bibitem{77a1eBrKhVo}
{V.B.Braginsky, Yu.I.Vorontsov, F.Ya.Khalili},
\newblock Sov.\,Phys.\,JETP {\bf 46}, 705 (1977).

\bibitem{Caves1981}
{C.M.Caves},
\newblock Physical Review D {\bf 23}, 1693 (1981).

\bibitem{02a1KiLeMaThVy}
{H.J.Kimble, Yu.Levin, A.B.Matsko, K.S.Thorne and S.P.Vyatchanin},
\newblock Physical Review D {\bf 65}, 022002 (2001).

\bibitem{Harms2003}
{Jan Harms, Yanbei Chen, Simon Chelkowski, Alexander Franzen, Hennig Vahlbruch,
  Karsten Danzmann, and Roman Schnabel},
\newblock Physical Review D {\bf 68}, 042001 (2003).

\bibitem{Buonanno2003}
{A.Buonanno, Y.Chen},
\newblock Physical Review D {\bf 67}, 062002 (2003).

\bibitem{Unruh1982}
{W.G.Unruh},
\newblock in {\em Quantum Optics, Experimental Gravitation, and Measurement
  Theory}, edited by {P.Meystre and M.O.Scully}, page 647, Plenum Press, New
  York, 1982.

\bibitem{87a1eKh}
{F.Ya.Khalili},
\newblock Doklady Akademii Nauk {\bf 294}, 602 (1987),
\newblock (see also \cite{92BookBrKh}, Chapter 8).

\bibitem{JaekelReynaud1990}
{M.T.Jaekel and S.Reynaud},
\newblock Europhysics Letters {\bf 13}, 301 (1990).

\bibitem{Pace1993}
{A.F.Pace, M.J.Collett and D.F.Walls},
\newblock Physical Review A {\bf 47}, 3173 (1993).

\bibitem{96a2eVyMa}
{S.P.Vyatchanin and A.B.Matsko},
\newblock Sov. Phys. JETP {\bf 83}, 690 (1996).

\bibitem{Arcizet2006}
{O. Arcizet, T. Briant, A. Heidmann and M. Pinard},
\newblock Physical Review A {\bf 73}, 033819 (2006).

\bibitem{Purdue2002}
{P.Purdue, Y.Chen},
\newblock Physical Review D {\bf 66}, 122004 (2002).

\bibitem{Buonanno2004}
{A.Buonanno, Y.Chen},
\newblock Physical Review D {\bf 69}, 102004 (2004).

\bibitem{90a1BrKh}
{V. B. Braginsky, F. Ya. Khalili},
\newblock Physics Letters A {\bf 147}, 251 (1990).

\bibitem{00a1BrGoKhTh}
{V.B.Braginsky, M.L.Gorodetsky, F.Ya.Khalili and K.S.Thorne},
\newblock Physical Review D {\bf 61}, 4002 (2000).

\bibitem{Purdue2001}
{P.Purdue},
\newblock Physical Review D {\bf 66}, 022001 (2002).

\bibitem{96a1KhLe}
{F. Ya. Khalili, Yu. Levin},
\newblock Physical Review D {\bf 54}, 4735 (1996).

\bibitem{Chen2002}
{Y.Chen},
\newblock Physical Review D {\bf 67}, 122004 (2003).

\bibitem{02a2Kh}
{F.Ya.Khalili},
\newblock arXive:gr-gc/0211088  (2002).

\bibitem{04a1Da}
{S.L.Danilishin},
\newblock Physical Review D {\bf 69}, 102003 (2004).

\bibitem{Byer1996}
{K-X.Sun, M.M.Fejer, E.Gustafson, D.Shoemaker, and R.L.Byer},
\newblock Physical Review Letters {\bf 76}, 3055 (1996).

\bibitem{Byer1999}
{P.Beyersdorf, M.M.Fejer, and R.L.Byer},
\newblock Optics Letters {\bf 24}, 1112 (1999).

\bibitem{Byer2000}
{S.Traeger, P.Beyersdorf, L.Goddard, E.Gustafson, M.M.Fejer, and R.L.Byer},
\newblock Optics Letters {\bf 25}, 722 (2000).

\bibitem{MSSDC2005}
{H.Mueller-Ebhardt, K.Somiya, R.Schnabel, K.Danzmann and Y.Chen},
\newblock {Signal-recycled Sagnac interferometer}, 2005,
\newblock {unpublished manuscript}.

\bibitem{Meers1988}
{B. J. Meers},
\newblock Physical Review D {\bf 38}, 2317 (1988).

\bibitem{Buonanno2001}
{A.Buonanno, Y.Chen},
\newblock Physical Review D {\bf 64}, 042006 (2001).

\bibitem{Buonanno2002}
{A.Buonanno, Y.Chen},
\newblock Physical Review D {\bf 65}, 042001 (2002).

\bibitem{MuellerEbhardt2009}
{Helge Mueller-Ebhardt, Henning Rehbein, Chao Li, Yasushi Mino, Kentaro Somiya,
  Roman Schnabel, Karsten Danzmann, and Yanbei Chen},
\newblock arXiv:0903.0798  (2009).

\bibitem{00a1JuBlZh}
{Ju, L, Blair, D.G., and Zhao, C},
\newblock Reports on Progress in Physics {\bf 63}(9), 1317 (2000).

\bibitem{99a1BrKh}
{V.B.Braginsky, F.Ya.Khalili},
\newblock Physics Letters A {\bf 257}, 241 (1999).

\bibitem{01a2Kh}
{F.Ya.Khalili},
\newblock Physics Letters A {\bf 288}, 251 (2001).

\bibitem{Rehbein2007}
{H.Rehbein, H.M\"uller-Ebhardt, K.Somiya, C.Li, R.Schnabel, K.Danzmann, and
  Y.Chen},
\newblock Physical Review D {\bf 76}, 062002 (2007).

\bibitem{97a1BrGoKh}
{V.B.Braginsky, M.L.Gorodetsky, F.Ya.Khalili},
\newblock Physics Letters A {\bf 232}, 340 (1997).

\bibitem{02a1Kh}
{F.Ya.Khalili},
\newblock Physics Letters A {\bf 298}, 308 (2002).

\bibitem{98a1BrGoKh}
{V.B.Braginsky, M.L.Gorodetsky, F.Ya.Khalili},
\newblock Physics Letters A {\bf 246}, 485 (1998).

\bibitem{03a1Kh}
{F.Ya.Khalili},
\newblock Physics Letters A {\bf 317}, 169 (2003).

\bibitem{06a1DaKh}
{S.L.Danilishin, F.Ya.Khalili},
\newblock Physical Review D {\bf 73}, 022002 (2006).

\bibitem{06a2Kh}
{F.Ya.Khalili},
\newblock Physical Review D {\bf 76}, 102002 (2007).

\end{thebibliography}

\end{document}